\documentclass[a4paper,11pt]{article}
\usepackage[DIV12]{typearea}
\usepackage{longtable}
\usepackage{booktabs}
\usepackage{amsmath}
\usepackage{amssymb}
\usepackage{bbm}
\usepackage[utf8]{inputenc}
\usepackage{pdflscape}
\usepackage{xspace}
\usepackage[caption=false]{subfig}
\usepackage{nicefrac}
\usepackage{graphicx}
\usepackage{cite}  
\usepackage{nccfoots}

\newcommand{\ket}[1]{\ensuremath|#1\rangle}
\newcommand{\rep}[1]{\ensuremath\boldsymbol{#1}}
\newcommand{\crep}[1]{\ensuremath\bar{\boldsymbol{#1}}}
\newcommand{\Z}[1]{\ensuremath{\mathbbm{Z}_{#1}}} % z_N ->\Z{N}
\newcommand{\SO}[1]{\ensuremath{\mathrm{SO}(#1)}}

\newcommand{\SL}[1]{\ensuremath{\mathrm{SL}(#1)}}
\newcommand{\U}[1]{\ensuremath{\mathrm{U}(#1)}}
\newcommand{\E}[1]{\ensuremath{\mathrm{E}_{#1}}}
\newcommand{\I}{\mathrm{i}}
\newcommand{\Id}{\mathbbm{1}}

\newcommand{\CP}{\ensuremath{\mathcal{CP}}\xspace}

\newcommand{\nphantom}[1]{\sbox0{#1}\hspace{-\the\wd0}}

\usepackage{xcolor}

\definecolor{darkgreen}{HTML}{109930}

\advance \headheight by 3.0truept       % for 12pt mandatory...
\usepackage[pdftex]{hyperref}

\hypersetup{
    pdftitle = {A String Theory of Flavor and CP},
    pdfauthor = {Baur, Nilles, Trautner, Vaudrevange}
}
 
\begin{document}

\begin{titlepage}

%\vspace*{-3.0cm}
\begin{flushright}
\normalsize{TUM-HEP 1218/19}
\end{flushright}

\vspace*{1.0cm}

\begin{center}
{\Large\textbf{A String Theory of Flavor and \boldmath $\mathcal{CP}$\unboldmath}}

\vspace{1cm}

\textbf{Alexander Baur}$^{a}$,
\textbf{Hans~Peter~Nilles}$^{b}$, \textbf{Andreas Trautner}$^{c}$, \\ and \textbf{Patrick~K.S.~Vaudrevange}$^{a}$
\Footnote{*}{%
\href{mailto:alexander.baur@tum.de;nilles@th.physik.uni-bonn.de;trautner@mpi-hd.mpg.de;patrick.vaudrevange@tum.de}{\tt Electronic addresses} 
}
\\[5mm]
$^a$~Physik Department T75, Technische Universit\"at M\"unchen,\\
James-Franck-Stra\ss e 1, 85748 Garching, Germany\\
\vspace{2mm}
$^b$~Bethe Center for Theoretical Physics and
Physikalisches Institut der Universit\"at Bonn,\\
Nussallee 12, 53115 Bonn, Germany\\
\vspace{2mm}
$^{c}$ Max-Planck-Institut f\"ur Kernphysik, \\ 
Saupfercheckweg 1, 69117 Heidelberg, Germany
\end{center}

\vspace{1cm}

\vspace*{1.0cm}

\begin{abstract}
Modular transformations of string theory (including the well-known stringy dualities) play a 
crucial role in the discussion of discrete flavor symmetries in the Standard Model. They are at the 
origin of \CP-transformations and provide a unification of \CP with traditional flavor symmetries. 
Here, we present a novel, fully systematic method to reliably compute the unified flavor symmetry 
of the low-energy effective theory, including enhancements from the modular transformations of 
string theory. The unified flavor group is non-universal in moduli space and exhibits the 
phenomenon of ``Local Flavor Unification'' where different sectors of the theory can be subject to 
different flavor structures.
\end{abstract}

\end{titlepage}

\newpage

%%%%%%%%%%%%%%%%%%%%%%%%%%%%%%%%%%%%%%%%%%%%%%%%%%%%%%%%%%%%%%%%%%%%%%%%%%%%%%%%%%%%%%%%%%%%%%%%%%%%%%%%%%%%%%%%%%%%%%%%%%%%%%%%%%%%%%%%%
\section{Introduction}

The origin of flavor remains one of the most challenging questions in the Standard Model (SM) of 
particle physics. String theory, as a consistent ultra-violet completion of the SM, can provide 
some useful ideas to attack this puzzle. Previous discussions of the origin of flavor symmetry in 
string theory~\cite{Kobayashi:2006wq,Nilles:2012cy,Beye:2014nxa} relied on some ``guesswork'' based 
on the geometry of compactified space and properties of string selection rules. While this led to 
models with appealing discrete flavor symmetries, it typically did not address the origin of \CP. A 
first step to include \CP was made in ref.~\cite{Nilles:2018wex}, where a \CP candidate was 
identified as an outer automorphism of the traditional flavor group. This provided a string theory 
origin of the general mechanism of group theoretical \CP violation discussed earlier 
in~\cite{Chen:2009gf,Feruglio:2012cw,Holthausen:2012dk,Chen:2014tpa,Trautner:2016ezn,Chen:2019iup}. 
Still, a comprehensive picture of the origin of flavor and \CP remained illusive: A priori, it is 
not clear whether the interpretation of the geometry of compact dimensions and string theory (space 
group) selection rules~\cite{Hamidi:1986vh,Dixon:1986qv,Ramos-Sanchez:2018edc} gives the complete 
set of symmetries. A more general mechanism is needed to clarify the situation.

\begin{figure}[t]
\centering
\includegraphics[width=0.8\linewidth]{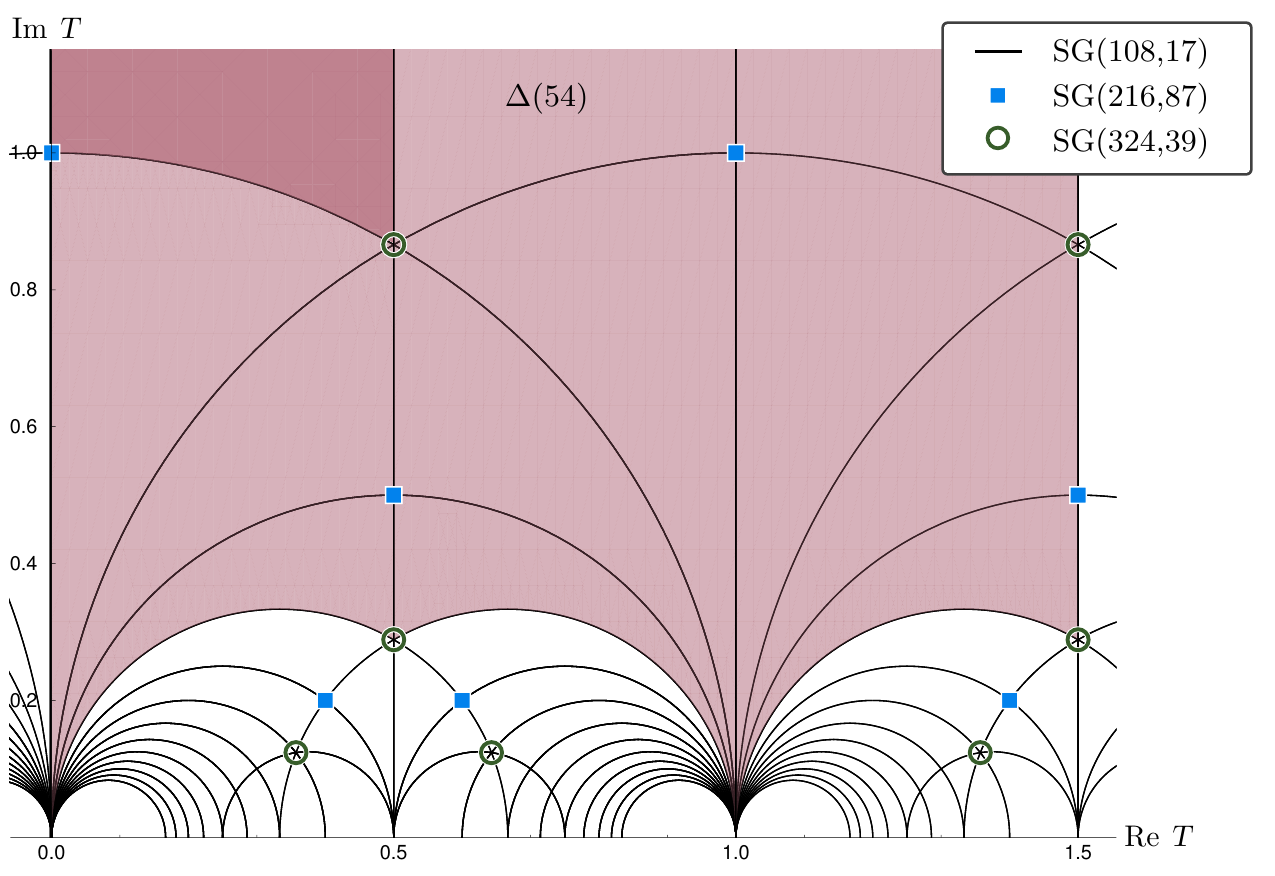}
\caption{\label{fig:modular_space1}
The moduli space of the K\"ahler modulus $T$ for the two-dimensional $\Z{3}$ orbifold. The 
traditional flavor symmetry $\Delta(54)$ is conserved at all points universally. If the K\"ahler 
modulus $T$ is stabilized on the indicated lines or points, the traditional flavor symmetry is 
enhanced by those modular transformations that leave the corresponding lines or points invariant. 
Importantly, these enhancements correspond to outer automorphisms of the traditional flavor symmetry.}
\end{figure}

In the present paper we shall present such a general mechanism. It is based on the consideration of 
outer automorphisms of the Narain lattice and the Narain space 
group~\cite{Narain:1985jj,Narain:1986am,Narain:1986qm}. The full set of symmetries is determined by 
the properties of the Narain space group~\cite{Baur:2019kwi}. In this way, the Narain space group 
encodes all the information from the string theory models under consideration, providing a unified 
description of the traditional flavor symmetry with \CP  (or \CP-like) transformations as its outer 
automorphisms~\cite{Nilles:2018wex}. Apart from the traditional flavor symmetries 
discussed so far, the full flavor symmetry uncovered in our approach also includes duality 
(modular) transformations that exchange winding and momentum states~\cite{Giveon:1988tt} and act 
nontrivially on the twisted states of string
theory~\cite{Lauer:1989ax,Lauer:1990tm,
Lerche:1989cs,Chun:1989se,Ferrara:1989bc,Ferrara:1989qb}.
This provides
a new perspective on the theory of flavor and \CP which was already outlined in our earlier 
paper~\cite{Baur:2019kwi}. The main results of the new scheme include: 
\begin{itemize}
\item The traditional flavor symmetries are only one part of this picture (these are the symmetries 
that are universal in the moduli space of string theory).
\item Modular (including duality) transformations of string theory are new ingredients of the full 
flavor structure. At some specific lower-dimensional regions (e.g. points or lines) in moduli 
space, the modular transformations become symmetries and lead to an enhancement of the flavor 
group, as illustrated in  figure~\ref{fig:modular_space1}.
\item \CP (or \CP-like) transformations are shown to be part of these modular 
transformations~\cite{Baur:2019kwi}. \CP is an exact symmetry only within the self-dual regions and 
spontaneously broken otherwise\footnote{%
This observation was reported in our earlier paper~\cite{Baur:2019kwi} and was subsequently 
incorporated in an explicit bottom-up flavor construction in~\cite{Novichkov:2019sqv}.}.
The modular enhanced flavor structure, therefore, leads to a unification of traditional flavor 
symmetries and \CP. 
\item The full unified flavor group is non-universal in moduli space, while the traditional flavor 
group is its subgroup that is preserved universally in moduli-space.
\item The non-universality of the flavor symmetry allows for different flavor structures
in different sectors of the theory. This allows the implementation of the significant difference of 
flavor structure in the quark and lepton sector of the Standard Model.
\end{itemize}

The paper is organized as follows. In section~\ref{sec:NarainAutomorphisms} we shall introduce the 
Narain lattice, the Narain space group, and its outer automorphisms. In section~\ref{sec:ModularSymmetry} 
we present technicalities of modular (and duality) transformations in two compact dimensions. The 
connections between modular transformations and flavor symmetries are subject of 
section~\ref{sec:ModularVSFlavor}. We shall discuss the enhancement of the traditional flavor 
symmetry and classify the possible enhanced symmetry groups in the two-dimensional case. In 
section~\ref{sec:Z3NarainFlavor} we consider the modular transformations of the two-dimensional 
$\Z{3}$ orbifold and classify all outer automorphisms of the corresponding $\Z{3}$ Narain space 
group. In section~\ref{sec:FlavorLandscape} we present the landscape of the enhanced flavor 
symmetries of the $\Z{3}$ orbifold. The traditional flavor symmetry is $\Delta(54)$ and it is 
universal in moduli space. On fixed lines and circles (of modular transformations) we find an 
enhancement of $\Delta(54)$ to $\mathrm{SG}(108, 17)$ which includes a \CP-like transformation 
as an exact symmetry. There are several different possibilities for such an 
enhancement, corresponding to several different ways of enhancing $\Delta(54)$ by different 
$\Z{2}$ subgroups of $\mathrm{S_4}$ -- the group of outer automorphisms of $\Delta(54)$. 
The location of the fixed straight lines and circles is shown in figure~\ref{fig:modular_space1}. 
The enhancements combine at points where two lines meet and the flavor group is enhanced to 
$\mathrm{SG}(216, 87)$. The maximal enhancement to $\mathrm{SG}(324, 39)$ is obtained at points 
where three lines meet. In section~\ref{sec:conclusions} we present the lessons from string theory 
for flavor model building. We shall compare the string theory point of view with previous bottom-up 
attempts in flavor model building that implement modular symmetries and \CP. More technical details 
of the construction are relegated to the appendices. Sections~\ref{sec:NarainAutomorphisms}-\ref{sec:Z3NarainFlavor} 
are rather technical and may be skipped in a first reading. The main results of the paper can be 
appreciated by looking at figures~\ref{fig:modular_space1} and \ref{fig:modular_space2} and reading 
sections~\ref{sec:FlavorLandscape} and~\ref{sec:conclusions}.

\section{Outer automorphisms of the Narain space group}
\label{sec:NarainAutomorphisms}
\subsection[Narain lattice]{\boldmath Narain lattice\unboldmath}

A toroidal compactification of $D$ bosonic string coordinates $y$ can be described most 
conveniently in the Narain formulation, where $y$ is separated into $D$ right- and $D$ left-moving 
degrees of freedom $y_\mathrm{R}$ and $y_\mathrm{L}$. Combined into a $2D$-dimensional Narain 
coordinate $Y$ this reads
\begin{equation}\label{eqn:CoordinatesYAndTildeY}
\begin{pmatrix}y\\\tilde{y}\end{pmatrix}~=~\frac{1}{\sqrt{2}}\begin{pmatrix}\Id&\Id\\-\Id&\Id\end{pmatrix}\,\begin{pmatrix}y_\mathrm{R}\\y_\mathrm{L}\end{pmatrix}\; \quad\mathrm{and}\quad Y ~:=~ \left(\begin{array}{c}y_\mathrm{R}\\y_\mathrm{L}\end{array}\right)\;,
\end{equation}
where $\tilde{y}$ denotes the T-dual coordinate of $y$. Then, $Y$ is compactified on a 
$2D$-dimensional torus that is parameterized by a Narain vielbein matrix $E$, i.e.\
\begin{equation}\label{eq:NarainTorusCompactification}
Y ~\sim~ Y + E\,\hat{N}\;,\quad\mathrm{where}\quad \hat{N} ~=~ \left(\begin{array}{c}n\\m\end{array}\right) ~\in~\Z{}^{2D}\;.
\end{equation}
Here, $\hat{N}\in\Z{}^{2D}$ contains the string's winding and Kaluza-Klein quantum numbers $n$ and 
$m$, respectively. To render the world-sheet theory modular invariant the Narain vielbein $E$ 
has to span an even self-dual lattice $\Gamma=\{E\,\hat{N} ~|~ \hat{N}\in\Z{}^{2D}\}$ with metric 
$\eta$ of signature $(D,D)$. Consequently, one can always choose $E$ such that 
\begin{equation}\label{eq:NarainConditionOnNarainVielbeinE}
E^\mathrm{T} \eta\, E ~=~ \hat\eta\;,  \quad\mathrm{where}\quad  \eta ~:=~ \begin{pmatrix} -\Id & 0\\ 0 & \Id\end{pmatrix}  \quad\mathrm{and}\quad  \hat\eta ~:=~ \begin{pmatrix} 0 & \Id\\ \Id & 0\end{pmatrix}\;,
\end{equation}
and the Narain vielbein can be parameterized as
\begin{equation}\label{eq:VielbeinMatrixE}
E ~:=~ \dfrac{1}{\sqrt{2}}\, \begin{pmatrix}
\dfrac{e^{-\mathrm{T}}}{\sqrt{\alpha'}}\,(G-B)&          -\sqrt{\alpha'}\,e^{-\mathrm{T}}\\[0.4cm]
\dfrac{e^{-\mathrm{T}}}{\sqrt{\alpha'}}\,(G+B)&\phantom{-}\sqrt{\alpha'}\,e^{-\mathrm{T}}\end{pmatrix}\;.
\end{equation}
In this definition of the Narain vielbein, $e$ denotes the vielbein of the $D$-dimensional 
geometrical torus $\mathbbm{T}^D$ with metric $G:=e^\mathrm{T}e$, $e^{-\mathrm{T}}$ corresponds to the 
inverse transposed matrix of $e$, $B$ is the anti-symmetric background $B$-field ($B=-B^\mathrm{T}$), 
and $\alpha'$ is called the Regge slope. Then, the scalar product of two Narain lattice vectors 
$\lambda_i = E\,\hat{N}_i \in \Gamma$ with $\hat{N}_i \in\Z{}^{2D}$ for $i\in\{1,2\}$ can be 
evaluated using eq.~\eqref{eq:NarainConditionOnNarainVielbeinE},
\begin{equation}\label{eq:NarainScalarProduct}
\lambda_1^\mathrm{T} \eta\, \lambda_2 ~=~ \hat{N}_1^\mathrm{T} E^\mathrm{T}\eta\,E\,\hat{N}_2 ~=~ \hat{N}_1^\mathrm{T} \hat\eta\,\hat{N}_2 ~=~ n_1^\mathrm{T}m_2 + m_1^\mathrm{T}n_2 ~\in~ \Z{}\;.
\end{equation}

As a remark, the Narain vielbein $E$ can be extended to the heterotic string by including Wilson 
line background fields that act on the 16 extra left-moving bosonic gauge degrees of freedom of the 
heterotic string, see e.g.~\cite{GrootNibbelink:2017usl}.

\subsection[Narain space group]{\boldmath Narain space group\unboldmath}
\label{sec:NarainSpaceGroup}

Next, we generalize the Narain lattice construction to $\Z{K}$ orbifolds in the Narain formulation. 
In this case, the toroidal compactification from eq.~\eqref{eq:NarainTorusCompactification} is 
extended by the action of a $\Z{K}$ Narain twist $\Theta$, i.e.
\begin{equation}\label{eq:NarainOrbifold}
Y ~\sim~ \Theta^k\,Y + E\,\hat{N}\;, \quad\mathrm{where}\quad \Theta ~=~ \left(\begin{array}{cc}\theta_\mathrm{R} & 0 \\0 & \theta_\mathrm{L}\end{array}\right) \quad\mathrm{and}\quad \Theta^K ~=~ \Id\;.
\end{equation}
The integer $k\in\{0,\ldots,K-1\}$ enumerates the twisted sectors. We have to demand the 
block-structure of the Narain twist $\Theta$ in eq.~\eqref{eq:NarainOrbifold} such that $\Theta$ 
cannot mix left- and right-moving modes of the string. In addition, the matrices $\theta_\mathrm{R}$ 
and $\theta_\mathrm{L}$ must be orthogonal in order to leave the string's mass invariant (c.f.\ 
eqs.~\eqref{eq:UntwistedStringMass} and \eqref{eqn:VertexOperatorZKOrbifoldInvariant} in 
appendix~\ref{sec:BulkVertexOperators}). For supersymmetric orbifolds it is moreover necessary that 
$\theta_\mathrm{R}\in\SO{D}$. We define a $\Z{K}$ Narain space group 
$S_\mathrm{Narain}$~\cite{GrootNibbelink:2017usl} as the multiplicative closure of a finite list of 
generators\footnote{%
The name ``$\Z{K}$ Narain space group'' refers to the $K$-fold twist $\Theta$ in 
$S_\mathrm{Narain}$ and should not indicate that the group is Abelian. By definition of 
$S_\mathrm{Narain}$, it is clear that it is actually non-Abelian, and, moreover, non-compact.}
\begin{equation}
S_\mathrm{Narain} ~:=~ \langle \;(\Theta, 0) \;\;,\;\; (\Id, E_i) \;\;\mathrm{for}\;\; i\in\{1,\ldots,2D\}\; \rangle\;,
\end{equation}
where the vector $E_i$ corresponds to the $i$-th column of the vielbein matrix $E$, and we restrict 
ourselves to $\Z{K}$ Narain space groups without roto-translations.\footnote{%
A $\Z{K}$ roto-translation would be generated by $(\Theta, V)$, 
where $V \not\in \Gamma$ but $(\Theta, V)^K = (\Id, \lambda)$ with $\lambda \in \Gamma$.} 
Then, a general $\Z{K}$ Narain space group element $g$ reads
\begin{equation}
g ~=~ (\Theta^k, E\,\hat{N}) ~\in~ S_\mathrm{Narain} \;\;\;\mathrm{where}\;\;\; k\in\{0,\ldots,K-1\} \;\;\;\mathrm{and}\;\;\; \hat{N}\in\Z{}^{2D}\;,
\end{equation}
and acts on the Narain coordinates $Y$ as $Y \mapsto g\, Y = \Theta^k\,Y + E\,\hat{N}$, see 
eq.~\eqref{eq:NarainOrbifold}. The Narain space group has to close under multiplication, 
especially $(\Theta, 0)\, (\Id, E\,\hat{N})\, (\Theta^{-1}, 0) = (\Id, \Theta\,E\,\hat{N})$ has to 
be an element of the Narain space group for all $\hat{N}\in\Z{}^{2D}$. Hence, $\Theta$ has to be an 
outer automorphism of the Narain lattice, $\Theta\,\Gamma = \Gamma$. Importantly, the 
Narain space group gives a natural framework to discuss the various classes of closed strings on 
Narain orbifolds, as briefly reviewed in appendix~\ref{app:StringStates}. Simultaneously it 
yields the discrete symmetries of the string setting via its outer automorphisms, as discussed in 
sections~\ref{sec:AutomorphismsNarainLattice} and~\ref{sec:AutomorphismsNarainSpaceGroup}.

Finally, by conjugation with the Narain vielbein $(E,0)$ we change the basis to the so-called 
lattice basis. To highlight this, all quantities in the lattice basis are written with a hat. For 
example, for each $g \in S_\mathrm{Narain}$ we define\footnote{To highlight whenever we work in the 
lattice basis, we also write $\hat{S}_\mathrm{Narain}$ for the Narain space group, even though 
$\hat{S}_\mathrm{Narain}\cong{S}_\mathrm{Narain}$ are, of course, one and the same group.}
\begin{equation}\label{eq:HatTheta}
\hat{g}~:=~(E^{-1},0)\,(\Theta^k,E\,\hat{N})\,(E,0)~=~(\hat\Theta^k,\hat{N})~\in~\hat{S}_\mathrm{Narain}\;,
\end{equation}
and the Narain twist in the lattice basis is defined as $\hat\Theta := E^{-1}\Theta\,E$, where 
$\hat\Theta \in \mathrm{GL}(2D,\Z{})$ follows automatically from the fact that $\hat\Theta$ is an 
outer automorphism of the Narain lattice. Depending on the choice of the Narain twist $\Theta$, the 
condition $\hat\Theta \in \mathrm{GL}(2D,\Z{})$ can freeze some of the free parameters (i.e.\ 
moduli) of the Narain vielbein $E$ to some special values.

Due to its general form in eq.~\eqref{eq:NarainOrbifold}, the twist in the lattice basis 
has to satisfy
\begin{equation}\label{eq:ConditionsOnHatTheta}
\hat\Theta^\mathrm{T} \hat\eta\,\hat\Theta ~=~ \hat\eta \quad\mathrm{and}\quad \hat\Theta^\mathrm{T} \mathcal{H}\,\hat\Theta ~=~ \mathcal{H}\;,
\end{equation}
where we have introduced the so-called generalized metric $\mathcal{H} := E^\mathrm{T}E$. Hence, 
the Narain twist $\hat\Theta$ leaves the Narain scalar product eq.~\eqref{eq:NarainScalarProduct}, 
as well as the generalized metric invariant.

In this work, we concentrate on so-called symmetric $\Z{K}$ orbifolds, meaning that we assume 
$\theta := \theta_\mathrm{R} = \theta_\mathrm{L}$, i.e.\ the Narain twist $\Theta$ acts 
left-right-symmetric.

\subsection{Outer automorphisms of the Narain lattice}
\label{sec:AutomorphismsNarainLattice}

A natural framework to understand the origin of modular transformations in string theory is the 
Narain lattice. In general, two lattices $\Gamma$ and $\Gamma'$ are identical if their vielbeins 
$E$ and $E'$ are related by a transformation $\hat\Sigma \in \mathrm{GL}(2D,\Z{})$, such 
that\footnote{We use the inverse matrix $\hat\Sigma^{-1}$ here for later convenience.} 
\begin{equation}\label{eqn:SymmetryTrafoOfNarainLattice}
E ~\mapsto~ E' ~=~ E\, \hat\Sigma^{-1} \quad\mathrm{for}\quad \hat\Sigma ~\in~ \mathrm{GL}(2D,\Z{})\;.
\end{equation}
If $\Gamma = \Gamma'$ is a Narain lattice we have to demand in addition that $\hat\Sigma$ leaves 
the Narain metric $\hat\eta$ invariant, $\hat\Sigma^\mathrm{T} \hat\eta\, \hat\Sigma ~=~ \hat\eta$.
This follows by using that both, $E$ and $E'$, satisfy eq.~\eqref{eq:NarainConditionOnNarainVielbeinE}. 
We use these conditions to define the group $\mathrm{O}_{\hat{\eta}}(D,D,\Z{})$ of outer 
automorphisms of the Narain lattice,
\begin{equation}\label{eqn:SymmetryGroupOfNarainLattice}
\mathrm{O}_{\hat{\eta}}(D,D,\Z{}) ~:=~ \big\langle ~\hat\Sigma~ \big| ~\hat\Sigma ~\in~ \mathrm{GL}(2D,\Z{}) \quad\mathrm{with}\quad \hat\Sigma^\mathrm{T} \hat\eta\, \hat\Sigma ~=~ \hat\eta ~\big\rangle\;.
\end{equation}
In section~\ref{sec:ModularSymmetry} below, we will analyze in more detail the group 
$\mathrm{O}_{\hat{\eta}}(D,D,\Z{})$ and its action on the moduli of the theory. We will then 
confirm that the outer automorphisms of the Narain lattice $\Gamma$ give rise to the modular 
transformations of this string setting.\footnote{Here, we ignore the continuous translational outer 
automorphisms of the Narain lattice as they correspond to additional $\U{1}$ symmetries, which will 
be broken to discrete subgroups in section~\ref{sec:AutomorphismsNarainSpaceGroup} after 
orbifolding}
Observe that, due to eq.~\eqref{eq:ConditionsOnHatTheta}, also the Narain twist $\hat\Theta$ must 
be an element of $\mathrm{O}_{\hat{\eta}}(D,D,\Z{})$, with the additional constraint 
$\hat\Theta^\mathrm{T} \mathcal{H}\,\hat\Theta = \mathcal{H}$. 

\subsection{Outer automorphisms of the Narain space group}
\label{sec:AutomorphismsNarainSpaceGroup}

After having discussed the outer automorphisms of the Narain lattice $\Gamma$, we continue with 
the outer automorphisms of the Narain space group $\hat{S}_\mathrm{Narain}$. They are of special 
interest as they yield the modular transformations as well as the flavor symmetries of the effective 
four-dimensional theory after orbifolding~\cite{Baur:2019kwi}. An outer automorphism of the Narain 
space group can be written as a transformation with
\begin{equation}\label{eq:OuterAutomorphismOfNarain}
\hat{h} ~:=~ (\hat\Sigma, \hat{T})~\not\in~\hat{S}_\mathrm{Narain}\;, 
\end{equation}
that acts as conjugation on $\hat{S}_\mathrm{Narain}$ and, by the definition of an automorphism, 
maps the Narain space group to itself. In detail, for \textit{each} element 
$(\hat\Theta^k,\hat{N}) \in \hat{S}_\mathrm{Narain}$ we have to ensure that 
\begin{equation}\label{eq:ConditionOnOuterAutomorphismOfNarain}
(\hat\Theta^k,\hat{N}) ~\stackrel{\hat{h}}{\mapsto}~ (\hat\Sigma,\hat{T})\; (\hat\Theta^k,\hat{N})\; (\hat\Sigma,\hat{T})^{-1} ~\stackrel{!}{\in}~ \hat{S}_\mathrm{Narain}\;.
\end{equation}
The action of such an outer automorphism on strings is discussed in 
appendix~\ref{app:TransformationsOfClosedStrings}. This Narain approach can be viewed as a stringy 
completion of a purely geometrical approach to identify flavor symmetries, for example in complete 
intersection Calabi-Yau manifolds~\cite{Lukas:2017vqp,Candelas:2017ive}. As discussed in 
ref.~\cite{Baur:2019kwi} one can find a set of generators of the group of outer automorphisms that 
is of the form
\begin{equation}
\left\{ (\hat\Sigma_1, 0),\; (\hat\Sigma_2, 0),\;\ldots,\; (\Id, \hat{T}_1),\; (\Id, \hat{T}_2),\;\ldots \right\}\;.
\end{equation}
In other words, the group of outer automorphisms can be generated by pure twists 
$(\hat\Sigma_i, 0)\not\in \hat{S}_\mathrm{Narain}$ and pure translations $(\Id, \hat{T}_j)\not\in \hat{S}_\mathrm{Narain}$. 
Roto-translations are not needed to generate the outer automorphism group.\footnote{However, 
this statement in general \textit{does not} hold in the case when the Narain space group 
$\hat{S}_\mathrm{Narain}$ itself \textit{has} roto-translations as generators.} Consequently, the 
translational part $\hat{T}_j$ of $(\Id, \hat{T}_j) \not\in \hat{S}_\mathrm{Narain}$ must be 
fractional, i.e.\ $\hat{T}_j\not\in\Z{}^{2D}$ and $0 \leq \hat{T}_j < 1$. Moreover, the twist 
$\hat\Sigma_i$ of $(\hat\Sigma_i, 0)\not\in \hat{S}_\mathrm{Narain}$ cannot be a rotation of the 
type $\Z{K}$ that has been used to construct the orbifold, i.e.\ $\hat\Sigma_i\neq \hat\Theta^\ell$ 
for $\ell\in\{0,\ldots,K-1\}$. Otherwise, $(\Id, \hat{T}_j)$ and $(\hat\Sigma_i, 0)$ would be inner 
automorphisms of the Narain space group, which is excluded by assumption.

The condition~\eqref{eq:ConditionOnOuterAutomorphismOfNarain} has important consequences for the 
special case of pure lattice translations, $(\Id, \hat{N}) \in \hat{S}_\mathrm{Narain}$, and 
purely rotational outer automorphisms, $\hat{h} = (\hat\Sigma, 0) \not\in \hat{S}_\mathrm{Narain}$. 
Namely, taking
\begin{equation}
(\Id, \hat{N}) ~\stackrel{\hat{h}}{\mapsto}~ (\hat\Sigma, 0)\; (\Id, \hat{N})\; (\hat\Sigma^{-1}, 0) ~=~ (\Id, \hat\Sigma\,\hat{N})~\stackrel{!}{\in}~ \hat{S}_\mathrm{Narain}\;,
\end{equation}
we find that $\hat{N} \stackrel{\hat{h}}{\mapsto} \hat\Sigma\,\hat{N}\stackrel{!}{\in}\Z{}^{2D}$, 
for any $\hat{N} \in \Z{}^{2D}$. Hence, a necessary condition for any twist 
$\hat\Sigma \in \mathrm{GL}(2D,\Z{})$ to be an outer automorphism $\hat{h} = (\hat\Sigma, 0)$ of 
the Narain space group is that $\hat\Sigma$ itself has to be an outer automorphism of the Narain 
lattice, that is
\begin{equation}
\hat\Sigma ~\in~ \mathrm{O}_{\hat{\eta}}(D,D,\Z{})\;.
\end{equation}

\section[Modular transformations]{\boldmath Modular transformations\unboldmath}
\label{sec:ModularSymmetry}

Let us now specialize to $D=2$ dimensions and analyze how transformations 
\mbox{$E \mapsto E'= E\, \hat\Sigma^{-1}$} with $\hat\Sigma \in \mathrm{O}_{\hat{\eta}}(2,2,\Z{})$ 
act on the K{\"a}hler ($T$) and complex structure modulus ($U$). Specifically, we will recapitulate 
how the modular transformations of the theory originate from the symmetries of the Narain lattice. 

In $D=2$ dimensions a general two-torus $\mathbbm{T}^2$ is parameterized by three real numbers 
(the lengths of the basis vectors $e_1$ and $e_2$ of the geometrical vielbein $e$ and their 
relative angle $\phi$). The strength of the anti-symmetric $B$-field is another free 
parameter. These four numbers can be combined into the so-called K{\"a}hler modulus\footnote{%
The K\"ahler modulus $T$ here should not be confused with the outer automorphism translations 
$\hat T$ above or the abstract $\SL{2,\Z{}}$ group element $\mathrm{T}$ below. We think it is 
always self-explanatory and clear from the context when we refer to which object.} 
$T$ and complex structure modulus $U$, which are given by
\begin{subequations} \label{eqn:KahlerandComplexStructure}
\begin{eqnarray}
T & = & T_1 + \I\, T_2 ~:=~ \frac{1}{\alpha'} \left(B_{12} + \I\, \sqrt{\det G}\right)\;,\label{eq:ParametersInTModulus}\\
U & = & U_1 + \I\, U_2 ~:=~ \frac{1}{G_{11}} \left(G_{12} + \I\, \sqrt{\det G}\right) ~=~ \frac{|e_2|}{|e_1|}\,\textrm{e}^{\I \phi}\;.\label{eq:ParametersInUModulus}
\end{eqnarray}
\end{subequations}
$T$ and $U$ describe all deformations of the $(2,2)$ Narain lattice. Consider now the matrices 
\begin{subequations}\label{eqn:ModularTrafosTandS}
\begin{eqnarray}
\hat{K}_\mathrm{S} ~:=~ \begin{pmatrix} 0 & \epsilon\\ \epsilon & 0\end{pmatrix} & \mathrm{and} & \hat{K}_\mathrm{T} ~:=~ \begin{pmatrix} \Id_2 & 0\\ -\epsilon & \Id_2\end{pmatrix}\;,   \quad\mathrm{with}\quad \epsilon ~:=~ \begin{pmatrix} 0 & 1\\ -1 & 0\end{pmatrix}\;,\\
\hat{C}_\mathrm{S} ~:=~ \begin{pmatrix} \epsilon & 0\\ 0 & \epsilon\end{pmatrix} & \mathrm{and} & \hat{C}_\mathrm{T} ~:=~ \begin{pmatrix} \gamma & 0\\ 0       & \gamma^{-T}\end{pmatrix}\;, \quad\mathrm{with}\quad \gamma ~:=~ \begin{pmatrix} 1 &-1\\  0 & 1\end{pmatrix}\;.
\end{eqnarray}
\end{subequations}
It is easily confirmed that $\hat{K}_{\mathrm{S, T}}$ and $\hat{C}_{\mathrm{S, T}}$ are elements of 
$\mathrm{O}_{\hat{\eta}}(2,2,\Z{})$, as defined in eq.~\eqref{eqn:SymmetryGroupOfNarainLattice}.
Futhermore, using the presentation
\begin{equation}
\mathrm{SL}(2,\Z{}) = \langle\; \mathrm{S}, \mathrm{T} ~|~ \mathrm{S}^4 ~=~ \Id \;,\; \mathrm{S}^2 ~=~ \left(\mathrm{S}\,\mathrm{T}\right)^3 \;\rangle\;,
\end{equation}
and noting that $\hat{K}_\mathrm{S}^2 = \hat{C}_\mathrm{S}^2=-\Id$, one confirms that 
$\hat{K}_{\mathrm{S, T}}$ and $\hat{C}_{\mathrm{S, T}}$ generate the modular group 
$\left[\mathrm{SL}(2,\Z{})_T\times\mathrm{SL}(2,\Z{})_U\right]/\Z2\subset\mathrm{O}_{\hat{\eta}}(2,2,\Z{})$. 
The $\Z{2}$ quotient here is generated by the abstract element 
$(\mathrm{S}_T)^2(\mathrm{S}_U)^2$ with $\mathrm{S}_T\in\mathrm{SL}(2,\Z{})_T$ and 
$\mathrm{S}_U\in\mathrm{SL}(2,\Z{})_U$, corresponding to the elements $\hat{K}_\mathrm{S}$ and $\hat{C}_\mathrm{S}$
of the representation~\eqref{eqn:ModularTrafosTandS}.
Finally, the set of generators of $\mathrm{O}_{\hat{\eta}}(2,2,\Z{})$ is completed by the mutually 
commuting $\Z2$ matrices\footnote{%
Note that the other factorized duality $\hat{M}'$ is also part of $\mathrm{O}_{\hat{\eta}}(2,2,\Z{})$ 
through the following relation: \mbox{$\hat{M}' \,=\, \hat{K}_\mathrm{S}^3\,\hat{C}_\mathrm{S}\,\hat{M}$}.}
\begin{equation}\label{eqn:ModularTrafosMStar}
\hat\Sigma_* ~:=~\left(\begin{array}{cccc}
-1 & 0 &  0 & 0 \\
 0 & 1 &  0 & 0 \\
 0 & 0 & -1 & 0 \\
 0 & 0 &  0 & 1 
\end{array}\right) \quad\mathrm{and}\quad \hat{M} ~:=~ \left(\begin{array}{cccc}
 0 & 0 & 1 & 0 \\
 0 & 1 & 0 & 0 \\
 1 & 0 & 0 & 0 \\
 0 & 0 & 0 & 1 
\end{array}\right)\;.
\end{equation}
Altogether this establishes the structure of $\mathrm{O}_{\hat{\eta}}(2,2,\Z{})$ as
\begin{equation}\label{eq:ModularGroupAsSL2Zs}
 \mathrm{O}_{\hat{\eta}}(2,2,\Z{})\cong \left[\left(\mathrm{SL}(2,\Z{})_T\times\mathrm{SL}(2,\Z{})_U\right)\rtimes \left(\Z2\times\Z2\right)\right]/\Z2\;.
\end{equation}

Next, we compute the transformation properties of the moduli $T$ and $U$ under 
$\hat{K}_\mathrm{S, T}$, $\hat{C}_\mathrm{S, T}$, $\hat\Sigma_*$, and $\hat M$. It is convenient to 
use the generalized metric $\mathcal{H} = E^\mathrm{T} E$ for this. As the Narain vielbein depends 
on the moduli $E = E(T,U)$, cf.\ eq.~\eqref{eq:VielbeinMatrixE}, so does the generalized metric 
$\mathcal{H}=\mathcal{H}(T,U)$. Under a transformation~\eqref{eqn:SymmetryTrafoOfNarainLattice} 
$\mathcal{H}$ transforms as
\begin{equation}\label{eqn:SymmetryTrafoOfGeneralizedMetric}
\mathcal{H}(T,U) ~\stackrel{\hat\Sigma}{\longmapsto}~ \mathcal{H}(T',U') ~=~ \hat\Sigma^{-\mathrm{T}}\mathcal{H}(T,U)\hat\Sigma^{-1}\;.
\end{equation}
This equation can be used to read off the transformations of the moduli,\footnote{%
	As an alternative, one may use section 3.3 of ref.~\cite{GrootNibbelink:2017usl} with 
	$\widehat{M} \in\{(\hat{K}_\mathrm{S})^{-1}, (\hat{K}_\mathrm{T})^{-1}, (\hat{C}_\mathrm{S})^{-1}, (\hat{C}_\mathrm{T})^{-1}, \hat\Sigma_*, \hat{M}\}$, 
	where we use the inverse matrices due to our altering defintion of the action in eq.~\eqref{eqn:SymmetryTrafoOfNarainLattice}.}
\begin{equation}\label{eqn:SymmetryTrafoOfModuli}
T ~\stackrel{\hat\Sigma}{\longmapsto}~ T' = T'(T,U) \quad\mathrm{and}\quad U ~\stackrel{\hat\Sigma}{\longmapsto}~ U' = U'(T,U)\;.
\end{equation}
Since $\pm\hat\Sigma$ both yield exactly the same transformation of the moduli in 
eq.~\eqref{eqn:SymmetryTrafoOfGeneralizedMetric}, each factor of $\mathrm{SL}(2,\Z{})$ acts only as 
$\mathrm{PSL}(2,\Z{})$ on the moduli. We find that $\hat{K}_\mathrm{S}$ and $\hat{K}_\mathrm{T}$ 
induce the transformations
\begin{subequations}
\begin{align}
\hat{K}_\mathrm{S} & ~:~ T ~\mapsto~ -\frac{1}{T}\;, & U ~&\mapsto~ U\;,& \\
\hat{K}_\mathrm{T} & ~:~ T ~\mapsto~ T + 1\;,        & U ~&\mapsto~ U\;,&
\end{align}
\end{subequations}
as expected for the modular group $\mathrm{PSL}(2,\Z{})_T$ of the K{\"a}hler modulus $T$. 
Furthermore, $\hat{C}_\mathrm{S}$ and $\hat{C}_\mathrm{T}$ generate the transformations
\begin{subequations}\label{eq:ComplexStructureTrafo}
\begin{align}
\hat{C}_\mathrm{S} & ~:~ T ~\mapsto~ T\;,&  U ~&\mapsto~ -\frac{1}{U}\;,&\\ 
\hat{C}_\mathrm{T} & ~:~ T ~\mapsto~ T\;,&  U ~&\mapsto~ U + 1\;,&
\end{align}
\end{subequations}
giving rise to the modular group $\mathrm{PSL}(2,\Z{})_U$ of the complex structure modulus $U$. 
Finally, $\hat\Sigma_*$ reflects the real parts of both, $T$ and $U$,
\begin{align}
\hat\Sigma_* & ~:~ T ~\mapsto~ -\bar{T}\;,& U ~&\mapsto~ -\bar{U}\;,&
\end{align}
while $\hat{M}$ interchanges the K{\"a}hler and complex structure modulus
\begin{align}
\hat{M} & ~:~ T ~\mapsto~ U\;,& U ~&\mapsto~ T \;,&
\end{align}
see e.g.\ ref.~\cite{Lerche:1989cs}. We will see later that $\hat\Sigma_*$ is related to \CP or 
\CP-like transformations~\cite{Baur:2019kwi}, while $\hat{M}$ induces the so-called mirror symmetry.

\newpage

\section[Connection between modular transformations and flavor symmetries]{\boldmath Connection between modular transformations and flavor symmetries\unboldmath}
\label{sec:ModularVSFlavor}

Modular transformations and flavor symmetries are closely connected in string 
theory~\cite{Baur:2019kwi}. We distinguish them here by their action on the moduli $T$ and $U$ and 
their breaking-behavior under non-vanishing vacuum expectation values (VEVs) of $\langle T\rangle$ 
and $\langle U\rangle$. 

As discussed in section~\ref{sec:NarainSpaceGroup}, orbifolding can freeze some of the moduli 
that parameterize the Narain vielbein $E$ to special values. Let us denote these frozen moduli 
collectively by $M_\mathrm{fix}$ and the unfrozen ones by $M$. For example, as we will see in 
detail in section~\ref{sec:Z3NarainFlavor}, for a symmetric $\Z{3}$ orbifold in $D=2$ 
dimensions the complex structure modulus is frozen, $M_\mathrm{fix} = U = \exp\left(2\pi\I/3\right)$, 
while the K\"ahler modulus remains unconstrained, $M=T$.

Now, let us consider an outer automorphism of the Narain space group, 
eq.~\eqref{eq:ConditionOnOuterAutomorphismOfNarain} and denote it by 
$\hat{h}=(\hat\Sigma, \hat{T}) \not\in \hat{S}_\mathrm{Narain}$, where 
$\hat\Sigma \in \mathrm{O}_{\hat{\eta}}(D,D,\Z{})$. Then, we can distinguish three cases of 
transformations:
\begin{enumerate}
\item[(1.)]
The traditional flavor symmetry. It is defined as the subgroup of the outer automorphisms of the 
Narain space group that remains unbroken at \textit{every} point in moduli space $\langle M\rangle$. 
All transformations $\hat{h}$ which trivially leave the moduli invariant,
\begin{equation}
M ~\stackrel{\hat{h}}{\longmapsto}~ M'(M) ~=~ M\;,
\end{equation}
belong to this class. In general, these transformations include all translational outer automorphisms 
\mbox{$\hat h=(\Id, \hat{T})\not\in \hat{S}_\mathrm{Narain}$} and the trivially acting element 
$\hat h=(-\Id,0)$. However, note that the latter element can also be an inner automorphism if it is 
taken to be part of the orbifold twist, in which case it does not appear as a flavor symmetry. 
Together, the translations $(\Id, \hat{T}_j)\not\in \hat{S}_\mathrm{Narain}$ and possibly the 
inversion $(-\Id, 0) \not\in\hat{S}_\mathrm{Narain}$ generate what we call the traditional flavor 
symmetry.

\item[(2.a)]
The modular transformations after orbifolding. These are given by those outer automorphisms 
$\hat{h}=(\hat\Sigma,\hat T)\not\in\hat{S}_\mathrm{Narain}$ that give rise to nontrivial modular 
transformations,
\begin{equation}
M ~\stackrel{\hat{h}}{\longmapsto}~ M'(M) ~\neq~ M\;,
\end{equation}
for generic values of the moduli $M$. Consequently, modular transformations are generally 
spontaneously broken by the VEVs of the moduli $\langle M\rangle$, i.e.\ at a generic point in 
moduli space. This may also include transformations involving complex conjugation and the mirror 
symmetry.

\item[(2.b)]
The unified flavor symmetry. This symmetry is a combination of the traditional flavor symmetry (1.) 
together with specific enhancements from the modular transformations (2.a) that depend on the 
location in moduli space. This happens due to the fact that some transformations from case (2.a) 
have fixed points, i.e.\ 
\begin{equation}\label{eq:FixedPointOfModuliTrafo}
M ~\stackrel{\hat{h}}{\longmapsto}~ M'(M) ~\stackrel{!}{=}~ M\;,
\end{equation}
for \textit{some} special values of the moduli $M$, even though in general $M'(M) ~\neq~ M$. If 
the VEVs of the moduli are stabilized precisely at these fixed points, then the corresponding outer 
automorphism $\hat{h}$ enhances the traditional flavor symmetry to build up the \textit{unified 
flavor symmetry}. For example, $M \mapsto M'(M) = -\nicefrac{1}{M}$ is a nontrivial modular 
transformation which has a fixed point at $M=\I$.

This enhancement may include modular transformations which involve complex conjugation or 
permutation of the moduli, which are generically related to \CP and \CP-like transformations or to 
so-called mirror symmetry, respectively. Hence, not all of these transformations are flavor 
symmetries in the traditional sense, for what reason we decided to call the resulting group the 
``unified flavor symmetry''. Depending on the localization in moduli space there can be various 
different ``unified flavor symmetries''. These unified flavor symmetries share the property that 
they are broken spontaneously to the traditional flavor symmetry, case (1.), once the moduli are 
deflected from any of their fixed points, eq.~\eqref{eq:FixedPointOfModuliTrafo}, to a generic 
point in moduli space.

\end{enumerate}

\subsection[Enhancements of the flavor symmetry]{\boldmath Enhancements of the flavor symmetry\unboldmath}
Let us discuss cases (1.) and (2.b) in more detail. As already remarked above, we exclude here 
transformations which are part of the orbifold twist $\hat\Theta$ and focus on the true outer automorphisms.

Focusing on case (2.b) we are looking for outer automorphisms of the Narain space group 
$\hat{h}=(\hat\Sigma, \hat{T})$ with $\hat\Sigma \in \mathrm{O}_{\hat{\eta}}(D,D,\Z{})$ that leave 
the moduli invariant \textit{only} at some special points but not at a generic point in moduli 
space.\footnote{The transformations that leave the moduli invariant at \textit{every} point in 
moduli space are classified as class (1.) above. Therefore, if they are not part of the orbifold 
twist, they are already included as part of the traditional flavor symmetry.} Hence, using 
eq.~\eqref{eqn:SymmetryTrafoOfGeneralizedMetric} we find the condition 
\begin{equation}\label{eq:GeneralizedMetricInvariant}
\mathcal{H}(M) ~\stackrel{\hat\Sigma}{\longmapsto}~ \mathcal{H}(M') ~=~ \hat\Sigma^{-\mathrm{T}}\mathcal{H}(M)\hat\Sigma^{-1} ~\stackrel{!}{=}~ \mathcal{H}(M)\;,
\end{equation}
which has to be solved for $M$ in order to identify values of the moduli with potentially enhanced 
symmetry. We can define $\Sigma := E\, \hat\Sigma\, E^{-1}$ and rewrite 
eq.~\eqref{eq:GeneralizedMetricInvariant} as
\begin{equation}\label{eq:TransformationsFromFlavor}
\Sigma^\mathrm{T} \Sigma ~\stackrel{!}{=}~ \Id\;.
\end{equation}
Moreover, combining this condition with $\Sigma^\mathrm{T} \eta\,\Sigma = \eta$ from 
eq.~\eqref{eqn:SymmetryGroupOfNarainLattice}, we have to demand
\begin{equation}\label{eq:StructureOfSigma}
\Sigma ~\stackrel{!}{=}~ \begin{pmatrix} \sigma_\mathrm{R}&0\\0&\sigma_\mathrm{L}\end{pmatrix} \quad\mathrm{where}\quad \sigma_\mathrm{R}, \sigma_\mathrm{L} ~\in~ \mathrm{O}(D) \quad\Rightarrow\quad \mathrm{det}(\sigma_\mathrm{R/L})=\pm 1\;,
\end{equation}
in analogy to the Narain twist $\Theta$ in eq.~\eqref{eq:NarainOrbifold}. We remark that this 
block-diagonal structure of orthogonal matrices in $\Sigma$ automatically ensures that the left- 
and right-moving masses of a general untwisted string are invariant under a transformation with 
$\Sigma$. All potential enhancements of the traditional flavor symmetry originate from outer 
automorphisms $(\hat\Sigma, 0) \not\in\hat{S}_\mathrm{Narain}$, where $\Sigma$ is of the form 
stated in eq.~\eqref{eq:StructureOfSigma}.

In the next section, we will make use of this block-diagonal structure of $\Sigma$ in $D=2$ 
dimensions to classify --- independent of the choice of orbifold twist --- all rotational outer 
automorphisms of the Narain space group that leave invariant the moduli at some special regions in 
moduli space. In contrast, the translational outer automorphisms depend on the chosen orbifold and, 
hence, they must be discussed case by case, see section~\ref{sec:Z3PureTranslations} for an example.

\subsection[Classifying the possible enhancements of flavor symmetries in D=2]{\boldmath Classifying the possible enhancements of flavor symmetries in $D=2$\unboldmath}
\label{sec:ClassesOfTwists}

We now specialize to the case of two extra dimensions compactified on a symmetric $\Z{K}$ orbifold 
with $K \neq 2$ and classify all possible enhancements of the traditional flavor 
symmetry.\footnote{%
We exclude the case $K=2$ of $\Z{2}$ orbifolds here, because 
for this case there are, in general, outer automorphisms which do not obey eq.~\eqref{eq:EqualDets}.
For example, for $K=2$ there can be outer automorphisms that act as a rotation on the right-mover 
and as a reflection on the left-mover.} 
All of them can be described by different $\Sigma$'s of the form stated in eq.~\eqref{eq:StructureOfSigma}. 
As discussed in ref.~\cite{Baur:2019kwi} the automorphism condition~\eqref{eq:ConditionOnOuterAutomorphismOfNarain} 
in $D=2$ restricts the determinants of $\sigma_\mathrm{R}$ and $\sigma_\mathrm{L}$. In more detail,
eq.~\eqref{eq:ConditionOnOuterAutomorphismOfNarain} implies that for all $k$ there is a $k'$ such 
that 
\begin{equation}\label{eq:ConditionsOnsigmaRL}
\sigma_\mathrm{R}\, \theta^k ~=~ \theta^{k'} \sigma_\mathrm{R} \quad\mathrm{and}\quad \sigma_\mathrm{L}\, \theta^k ~=~ \theta^{k'} \sigma_\mathrm{L}\;.
\end{equation}
The determinant of $\sigma_\mathrm{R} \in \mathrm{O}(2)$ is constrained to be $\pm 1$. Let us assume 
first that $\sigma_\mathrm{R}$ is a rotation and not a reflection, i.e.\ $\mathrm{det}(\sigma_\mathrm{R})=1$. 
Then, we use that in $D=2$ dimensions all rotations necessarily commute. Consequently, $k=k'$ in 
eq.~\eqref{eq:ConditionsOnsigmaRL} and we find
\begin{equation}
\sigma_\mathrm{L}\, \theta ~=~ \theta\, \sigma_\mathrm{L}\;.
\end{equation}
Moreover, by assumption we are considering the case $\theta \neq -\Id$ since $K\neq 2$. Hence, 
$\sigma_\mathrm{L}$ must be a rotation, too. Thus, in this case we get 
$\mathrm{det}(\sigma_\mathrm{R}) = \mathrm{det}(\sigma_\mathrm{L}) = +1$. Repeating these arguments for 
$\sigma_\mathrm{R}$ being a reflection, i.e.\ $\mathrm{det}(\sigma_\mathrm{R})=-1$, we obtain in general
\begin{equation}\label{eq:EqualDets}
\mathrm{det}(\sigma_\mathrm{R}) ~=~ \mathrm{det}(\sigma_\mathrm{L})\;.
\end{equation}
This restricts $\Sigma$ to four cases (now given in the lattice basis 
$\hat\Sigma = E^{-1} \Sigma\, E$): 
\begin{enumerate}
\item Symmetric rotations $\hat{S}_\mathrm{rot.}(\alpha)$ with $\sigma_\mathrm{R} = \sigma_\mathrm{L}$ and $\mathrm{det}(\sigma_\mathrm{R}) =+1$,
\item Symmetric reflections $\hat{S}_\mathrm{refl.}(\alpha)$ with $\sigma_\mathrm{R} = \sigma_\mathrm{L}$ and $\mathrm{det}(\sigma_\mathrm{R}) =-1$,
\item Asymmetric rotations $\hat{A}_\mathrm{rot.}(\alpha_\mathrm{R}, \alpha_\mathrm{L})$ with $\sigma_\mathrm{R} \neq \sigma_\mathrm{L}$ and $\mathrm{det}(\sigma_\mathrm{R}) = \mathrm{det}(\sigma_\mathrm{L}) = +1$,
\item Asymmetric reflections $\hat{A}_\mathrm{refl.}(\alpha_\mathrm{R}, \alpha_\mathrm{L})$ with $\sigma_\mathrm{R} \neq \sigma_\mathrm{L}$ and $\mathrm{det}(\sigma_\mathrm{R}) = \mathrm{det}(\sigma_\mathrm{L}) = -1$.
\end{enumerate}
Here, the symmetric transformations are parameterized by one angle $\alpha$ (being either the 
rotation angle or the angle of the reflection axis) and asymmetric transformations are 
parameterized by two angles $\alpha_\mathrm{R}$ and $\alpha_\mathrm{L}$ (being either the rotation 
angles or the angles of the reflection axes).

By the definition of outer automorphisms, the rotations must map the four-dimensional Narain 
lattice to itself. Thus, the order of a four-dimensional rotation (associated to 
$\hat{S}_\mathrm{rot.}(\alpha)$ or $\hat{A}_\mathrm{rot.}(\alpha_\mathrm{R}, \alpha_\mathrm{L})$) 
is restricted to the orders of the possibly allowed crystallographic rotations in four 
dimensions. These orders can easily be found from the Euler-$\phi$ 
function~\cite{schwarzenberger1980n}, and they are given by
\begin{equation}\label{eq:OrdersOfRotationIn4D}
\{1,2,3,4,5,6,8,10,12\}\;.
\end{equation}
Another condition is that the $4 \times 4$ matrix $\hat\Sigma$ must be integral in the lattice 
basis, c.f.\ eq.~\eqref{eqn:SymmetryGroupOfNarainLattice}. As this matrix is obtained from 
$\Sigma$ (given by the four cases above) via $\hat\Sigma = E^{-1} \Sigma\, E$, this integral 
condition is in general only fulfilled for special values of the moduli that parameterize the 
vielbein $E$ of the Narain lattice. This explains why the unified flavor symmetry -- 
originating from the outer automorphism group of the Narain space group -- can depend on the value 
of the moduli.

After having classified all outer automorphisms of the Narain space group, we also want to analyze how 
they act on both, untwisted and twisted string states. The transformation of untwisted strings can be 
computed easily as we show in appendix~\ref{sec:BulkVertexOperators}. However, in order to identify 
the actual flavor symmetry generated by the outer automorphisms, one also has to identify the transformation 
properties of twisted strings. We will proceed to do this for the example of a 
$\Z{3}$ orbifold in the next section.

One of the main explicit results of the present work then is a complete classification of all outer 
automorphisms $\hat{h}=(\hat\Sigma, \hat{T})\not\in \hat{S}_\mathrm{Narain}$ for the 
two-dimensional $\Z{3}$ orbifold presented in the next section.

\section[Outer automorphisms of the Z3 Narain space group]{\boldmath Outer automorphisms of the $\Z{3}$ Narain space group\unboldmath}
\label{sec:Z3NarainFlavor}

To be specific, we analyze the symmetric $\Z{3}$ orbifold in $D=2$ as our main example. We begin in 
section~\ref{sec:Z3NarainSpaceGroup} by defining the Narain space group of the symmetric $\Z{3}$ 
orbifold in $D=2$ dimensions. Then, in section~\ref{sec:Z3ModularTransformations}, we identify 
those modular transformations that remain unbroken after orbifolding and analyze the transformation 
properties of untwisted and twisted strings in section~\ref{sec:Z3ModularTransformationsOfStrings}. 
Finally, in section~\ref{sec:Z3AllFlavorSymmetries}, we classify the outer automorphisms into the 
above types (1.) and (2.b), i.e.\ into traditional and unified flavor symmetries, respectively.

\subsection[Z3 Narain space group]{\boldmath $\Z{3}$ Narain space group\unboldmath}
\label{sec:Z3NarainSpaceGroup}

In the lattice basis of the $(2,2)$ Narain formulation, the symmetric $\Z{3}$ twist $\hat\Theta$ 
reads (cf.\ section~\eqref{sec:NarainSpaceGroup})
\begin{equation}\label{eq:Z3TwistLatticeBasis}
\hat\Theta ~=~ \left(\begin{array}{cccc}
 0 & -1 & 0 &  0 \\
 1 & -1 & 0 &  0 \\
 0 &  0 &-1 & -1 \\
 0 &  0 & 1 &  0 
\end{array}\right) ~\in~ \mathrm{O}_{\hat{\eta}}(2,2,\Z{}) \quad\Leftrightarrow\quad \theta ~=~ \theta_\mathrm{R} ~=~ \theta_\mathrm{L} ~=~ \left(\begin{array}{cc}
-\frac{1}{2}        & -\frac{\sqrt{3}}{2} \\
 \frac{\sqrt{3}}{2} & -\frac{1}{2} \\
\end{array}\right)\;.
\end{equation}
It can be decomposed into the generators $\hat{C}_\mathrm{S}$ and $\hat{C}_\mathrm{T}$ of the 
modular group $\mathrm{SL}(2,\Z{})_U$ as 
\begin{equation}\label{eq:Z3TwistLatticeBasisDecomp}
\hat\Theta ~=~ \left(\hat{C}_\mathrm{S}\right)^3 \hat{C}_\mathrm{T} ~\in~ \mathrm{SL}(2,\Z{})_U\;.
\end{equation}
Consequently, we can use eq.~\eqref{eq:ComplexStructureTrafo} to show that the moduli $T$ and $U$ 
transform under the twist action~\eqref{eq:Z3TwistLatticeBasisDecomp} as
\begin{equation}
\hat\Theta  ~:~ T ~\mapsto~ T \quad,\quad U ~\mapsto~ -\frac{1}{U+1} \;.
\end{equation}
Therefore, $\hat\Theta$ is a symmetry of the Narain lattice $\Gamma$ spanned by the vielbein $E$ 
for an arbitrary value of the K{\"a}hler modulus $T$ but fixed complex structure modulus,
\begin{equation}\label{eq:Z3ComplexStructure}
U ~\stackrel{!}{=}~ -\frac{1}{U+1} \qquad \Leftrightarrow \qquad U ~=~ \exp\left(2\pi\I/3\right) \qquad \Rightarrow \qquad E ~=~ E(T)\;.
\end{equation}
Thus, the complex structure modulus $U$ is frozen at $\exp\left(2\pi\I/3\right)$ which corresponds 
to the case $R:=|e_1|=|e_2|$, enclosing an angle of 120$^\circ$. This might have been expected from 
geometrical considerations of the symmetric $\Z{3}$ orbifold, see eq.~\eqref{eq:ParametersInUModulus}. 
Hence, the metric $G$ and the $B$-field $B$ are fixed up to two free parameters: the overall radius 
$R$ and the parameter $b$ of the anti-symmetric $B$-field, i.e.
\begin{equation}
e~=~R\begin{pmatrix}1&-\frac{1}{2}\\[2pt]0 &\phantom{0} \frac{\sqrt{3}}{2}\end{pmatrix}\;,
\quad G~=~\frac{R^2}{2} \begin{pmatrix}2&-1\\[2pt] -1&2\end{pmatrix}\;, 
\quad\mathrm{and}\quad B~=~b\,\alpha' \begin{pmatrix}0&1\\[2pt]-1&0\end{pmatrix}\;.
\end{equation}
In this case, the K\"ahler modulus, eq.~\eqref{eq:ParametersInTModulus}, reads
\begin{equation}
T~=~b+\I\,\frac{\sqrt{3}}{2}\,r \;,
\end{equation}
where we have defined $r:=\nicefrac{R^2}{\alpha'}$.

\subsection[Modular transformations after orbifolding]{\boldmath Modular transformations after orbifolding\unboldmath}
\label{sec:Z3ModularTransformations}

Modular transformations were introduced in section~\ref{sec:ModularSymmetry} as outer automorphisms 
of the Narain lattice. Now, in order to remain unbroken after orbifolding, a modular transformation 
has to be an (outer) automorphism of the Narain space group $\hat{S}_\mathrm{Narain}$ as well, see 
case (2.a) in section~\ref{sec:ModularVSFlavor}. In the following we analyze which elements 
$\hat\Sigma$ of the modular group eq.~\eqref{eq:ModularGroupAsSL2Zs} of the general $(2,2)$ Narain 
lattice satisfy the condition that for each $k\in\{0,1,2\}$ there is a $k'\in\{0,1,2\}$ such that 
the $\Z{3}$ orbifold twist $\hat\Theta$ from eq.~\eqref{eq:Z3TwistLatticeBasis} satisfies the 
condition
\begin{equation}
\hat\Sigma\, \hat\Theta^k\, \hat\Sigma^{-1} ~=~ \hat\Theta^{k'}\;,
\end{equation}
which originates from eq.~\eqref{eq:ConditionOnOuterAutomorphismOfNarain}. In this case, 
$\hat\Sigma$ is an unbroken modular transformation after orbifolding.

We note that $\hat\Theta \in \mathrm{SL}(2,\Z{})_U$, see eq.~\eqref{eq:Z3TwistLatticeBasisDecomp}. 
Moreover, elements from $\mathrm{SL}(2,\Z{})_T$ commute with those from $\mathrm{SL}(2,\Z{})_U$.
It is therefore obvious that the generators $\hat{K}_\mathrm{S}$ and $\hat{K}_\mathrm{T}$ of 
$\mathrm{SL}(2,\Z{})_T$ commute with $\hat\Theta$,
\begin{equation}
\hat{K}_\mathrm{S}\, \hat\Theta\, \hat{K}_\mathrm{S}^{-1} ~=~ \hat\Theta \quad\mathrm{and}\quad \hat{K}_\mathrm{T}\, \hat\Theta\, \hat{K}_\mathrm{T}^{-1} ~=~ \hat\Theta\;.
\end{equation}
Hence, modular transformations from $\mathrm{SL}(2,\Z{})_T$ are automorphisms of the $\Z{3}$ Narain 
space group and, therefore, remain unbroken after orbifolding. Note that these transformations do 
not interchange the twisted sectors, i.e.\ a string with constructing element 
\mbox{$\hat{g}=(\hat\Theta^k,\hat{N})\in\hat{S}_\mathrm{Narain}$} from the $k$-th twisted sector is 
mapped by a modular transformation $\mathrm{SL}(2,\Z{})_T$ to a string from the same twisted 
sector, see eq.~\eqref{eq:TrafoOfStringState} in appendix~\ref{app:StringStates}.

On the other hand,
\begin{equation}
\hat{C}_\mathrm{S}\, \hat\Theta\, \hat{C}_\mathrm{S}^{-1} ~\neq~ \hat\Theta^{k'} \quad\mathrm{and}\quad \hat{C}_\mathrm{T}\, \hat\Theta\, \hat{C}_\mathrm{T}^{-1} ~\neq~ \hat\Theta^{k'}\;,
\end{equation}
for any $k' \in\{0,1,2\}$. 
Consequently, the generators of the modular group $\mathrm{SL}(2,\Z{})_U$ are not automorphisms of 
the $\Z{3}$ Narain space group -- in other words, $\mathrm{SL}(2,\Z{})_U$ is broken by the 
orbifold. This can also be understood in the following way: As we have seen in 
eq.~\eqref{eq:Z3ComplexStructure}, the complex structure modulus $U$ has to be frozen at 
$U = \exp\left(2\pi\I/3\right)$ for the symmetric $\Z{3}$ orbifold. Hence, any modular 
transformation that does not leave $U = \exp\left(2\pi\I/3\right)$ invariant must be broken. 
Indeed, there is a $\Z{6}$ subgroup of $\mathrm{SL}(2,\Z{})_U$, generated by 
$\hat{C}_\mathrm{S} \hat{C}_\mathrm{T}$, which leaves $U = \exp\left(2\pi\I/3\right)$ invariant. 
This $\Z{6}$ can be written as $\Z{3}\times\Z{2}$, where the $\Z{3}$ is generated by the orbifold 
twist $\hat\Theta = (\hat{C}_\mathrm{S})^3 \hat{C}_\mathrm{T}$ while the $\Z{2}$ is generated by 
$\hat{C}_\mathrm{S}^2$. Thus, the $\Z{3}$ factor is an inner, not an outer, automorphism of the 
Narain space group, while the $\Z{2}$ factor can be written as 
$\hat{C}_\mathrm{S}^2 = \hat{K}_\mathrm{S}^2$ and, hence, also appears from the unbroken 
modular group $\mathrm{SL}(2,\Z{})_T$. Consequently, the modular group $\mathrm{SL}(2,\Z{})_U$ does 
not contain any independent outer automorphisms of the $\Z{3}$ Narain space group and, therefore, does 
not contribute to the flavor symmetry.

In addition to elements of $\mathrm{SL}(2,\Z{})_{U}$ or $\mathrm{SL}(2,\Z{})_{T}$, we consider the 
$\CP$-like transformation $\hat{K}_*$ defined as
\begin{equation}
\hat{K}_* ~:=~ \hat{C}_\mathrm{S}\, \hat{C}_\mathrm{T}\, \hat{C}_\mathrm{S}\, \hat{\Sigma}_* ~=~ \left(\begin{array}{cccc}
 1 &  0 & 0 &  0 \\
 1 & -1 & 0 &  0 \\
 0 &  0 & 1 &  1 \\
 0 &  0 & 0 & -1 
\end{array}\right) \quad\mathrm{with}\quad \left(\hat{K}_*\right)^2 ~=~ \Id\;.
\end{equation}
This transformation is an outer automorphism of the $\Z{3}$ Narain space group
\begin{equation}\label{eqn:ActionOfSreflOnTwistedSectors}
\hat{K}_*\, \hat\Theta\, \hat{K}_*^{-1} ~=~ \hat\Theta^2\;,\\
\end{equation}
and hence remains unbroken after orbifolding. Very importantly, $\hat{K}_*$ interchanges strings 
from the first and second twisted sector, see eq.~\eqref{eq:TrafoOfStringState} in 
appendix~\ref{app:StringStates}. Consequently, eq.~\eqref{eqn:ActionOfSreflOnTwistedSectors} has 
significant consequences for the heterotic string: the heterotic string has 16 extra left-moving 
bosonic degrees of freedom $X^I$ for $I=1,\ldots,16$, which give rise to the 10D gauge symmetry, 
for example $\E{8}\times\E{8}$, and four right-moving complex world-sheet fermions $\Psi^a$ (in 
light-cone gauge). In order to promote the interchange of twisted sectors induced by $\hat{K}_*$ to 
an automorphism of the heterotic string, the action of $\hat{K}_*$ has to be extended to 
\begin{equation}
X^I ~\longmapsto~ -X^I \quad\mathrm{and}\quad \Psi^a ~\longmapsto~ \bar\Psi^a\;.
\end{equation}
Hence, the transformation $\hat{K}_*$ maps all gauge representations to their complex conjugates 
and interchanges left-chiral and right-chiral target-space fermions. Thus, it corresponds to a 
physical $\CP$ transformation of the gauge groups.\footnote{%
The corresponding (outer) automorphism of the gauged semi-simple Lie groups is unique and 
corresponds to the usual, most general physical \CP transformation~\cite{Grimus:1995zi}.
However, for the (discrete, possibly enhanced) flavor symmetry, it is possible that the corresponding transformation
``only'' acts as a \CP-like symmetry, which is exactly the mechanism of physical \CP violation
which we had already discussed in~\cite{Nilles:2018wex}.}
Finally, under a transformation $\hat{K}_*$ the 
moduli transform as
\begin{align}
\hat{K}_* & ~:~ T ~\mapsto~ -\bar{T}\;, &\hspace{-3cm} U ~\mapsto~ -\frac{\bar{U}}{1+\bar{U}} \;.
\end{align}
This leaves the special choice of moduli
\begin{equation}
T ~=~ \I\,T_2 \qquad\mathrm{and}\qquad U ~=~ \exp\left(2\pi\I/3\right)\;
\end{equation}
invariant for all values of $T_2\in\mathbbm{R}$. Since the complex structure modulus $U$ is frozen 
by the $\Z{3}$ orbifold to precisely this value according to eq.~\eqref{eq:Z3ComplexStructure}, we 
confirm that the transformation $\hat{K}_*$ is unbroken by the $\Z{3}$ orbifold \textit{as long as 
$T$ takes purely imaginary values}.

In conclusion, we find that the maximal modular group after orbifolding is generated by 
$(\hat{K}_\mathrm{S}, \hat{K}_\mathrm{T}, \hat{K}_*)$ and has the structure
\begin{equation}
\mathrm{SL}(2,\Z{})_T \rtimes \Z{2}\;,
\end{equation}
where the $\Z{2}$ factor acts as physical $\CP$ transformation, at least for the gauge sector. In 
fact, using the presentation~\cite{cohn2013algebra}
\begin{equation}\label{eq:GL2Z}
\mathrm{GL}(2,\Z{}) ~=~ \langle\; \mathrm{S}, \mathrm{T}, \mathrm{K} ~|~ \mathrm{S}^4 = \Id \;,\; \mathrm{S}^2 = \left(\mathrm{S}\,\mathrm{T}\right)^3 \;,\; \mathrm{K}^2 = \Id \;,\; \left(\mathrm{S}\,\mathrm{K}\right)^2 = \Id \;,\; \left(\mathrm{T}\,\mathrm{K}\right)^2 = \Id  \;\rangle\;,
\end{equation}
we find that our generators actually form a representation of the extended modular group 
$\mathrm{GL}(2,\Z{})\cong\mathrm{SL}(2,\Z{})\rtimes\Z{2}$. Very importantly, we will see next that 
all massless string states transform trivial under a certain subgroup of $\mathrm{GL}(2,\Z{})$ (i.e.\ 
under a congruence subgroup, where elements of $\mathrm{GL}(2,\Z{})$ are only defined mod 3). Hence, 
massless strings form representations of the finite modular group $\mathrm{GL}(2,3)$. Furthermore, 
it is clear that the \CP-like transformation $\hat{K}_*$ can be spontaneously broken,
if $\langle T\rangle$ is deflected away from a purely imaginary value.

\subsection[Modular transformations of strings]{\boldmath Modular transformations of strings\unboldmath}
\label{sec:Z3ModularTransformationsOfStrings}

Next, we discuss strings on the symmetric $\Z{3}$ orbifold in $D=2$ and their transformation 
properties under the modular group $\mathrm{GL}(2,\Z{})$ generated by $\hat{K}_\mathrm{S}$, 
$\hat{K}_\mathrm{T}$, and $\hat{K}_*$. A detailed description of strings on orbifolds is reviewed 
in appendix~\ref{app:StringStates}. In summary, there are two classes of strings on orbifolds, 
invariant under the $\Z{3}$ twist: First of all, there are untwisted strings 
$V(\hat{N})^\mathrm{orb.}$ with constructing elements 
$[g]=\{(\Id, E\,\hat{N}), (\Id, E\,\hat\Theta\,\hat{N}), (\Id, E\,\hat\Theta^2\,\hat{N})\} \subset S_{\mathrm{Narain}}$. 
Here, $\hat{N} = (n_1, n_2, m_1, m_2)^\mathrm{T} \in \Z{}^4$ parameterizes the Kaluza-Klein (KK) 
momentum $(m_1, m_2)$ and winding numbers $(n_1, n_2)$ of the untwisted string on the orbifold. At 
a generic point in $T$-moduli space, only the untwisted string with $\hat{N}=(0,0,0,0)^\mathrm{T}$ 
is massless. Yet, we are interested in the full tower of massive strings. It is convenient to group 
these untwisted strings into nine classes of strings $V^{(M,N)}$ for $M,N \in\{0,1,2\}$ according 
to their discrete KK and winding charges, which are defined modulo 3 as
\begin{equation}
(M, N) ~=~ (-m_1 + m_2, n_1 + n_2)\;.
\end{equation}
Second, there are twisted strings: We denote them by $(X,\,Y,\,Z)$ and $(\bar{X},\,\bar{Y},\,\bar{Z})$ 
for the three twisted strings localized at the three fixed points in the first and second twisted 
sector, respectively. Note that $(\bar{X},\,\bar{Y},\,\bar{Z})$ give rise to the right-chiral 
CPT-conjugates of $(X,\,Y,\,Z)$ needed to promote $(X,\,Y,\,Z)$ to complete left-chiral superfields.

\begin{table}[!t]
\begin{center}
\begin{tabular}{c|c|c|c}
\toprule
generator          & transformation           & transformation        & six-dimensional representation $\hat{\Sigma}_{\rep{6}}$\\ 
$\hat{\Sigma}$     & of                       & of charges $(M,N)$    & for twisted strings\\
                   & $T$-modulus              & for untwisted strings & $(X,\,Y,\,Z,\,\bar{X},\,\bar{Y},\,\bar{Z})$\\
\addlinespace
\midrule
\addlinespace
$\hat{K}_\mathrm{S}$ & $T~\mapsto~-\frac{1}{T}$ & $(M,N)\mapsto(N,-M)$       & $-\frac{\I}{\sqrt{3}}\begin{pmatrix} 1 & 1 & 1 & 0 & 0 & 0\\ 1 & \omega & \omega^2 & 0 & 0 & 0\\ 1 & \omega^2 & \omega & 0 & 0 & 0\\ 0 & 0 & 0 & -1 & -1 & -1\\ 0 & 0 & 0 & -1 & -\omega^2 & -\omega\\ 0 & 0 & 0 & -1 & -\omega & -\omega^2\end{pmatrix}$\\
\addlinespace
\midrule[0.15mm]
\addlinespace
$\hat{K}_\mathrm{T}$ & $T~\mapsto~T+1$          & $(M,N)\mapsto(M-N,N)$      & $\phantom{-\frac{\I}{\sqrt{3}}}\begin{pmatrix} \omega^2 & 0 & 0 & 0 & 0 & 0\\ 0 & 1 & 0 & 0 & 0 & 0\\ 0 & 0 & 1 & 0 & 0 & 0\\ 0 & 0 & 0 & \omega & 0 & 0\\ 0 & 0 & 0 & 0 & 1 & 0\\ 0 & 0 & 0 & 0 & 0 & 1\end{pmatrix}$\\
\addlinespace
\midrule[0.15mm]
\addlinespace
$\hat{K}_*$          & $T~\mapsto~-\bar{T}$     & $(M,N)\mapsto(M,-N)$       & $\phantom{-\frac{\I}{\sqrt{3}}}\begin{pmatrix} 0 & 0 & 0 & 1 & 0 & 0\\ 0 & 0 & 0 & 0 & 1 & 0\\ 0 & 0 & 0 & 0 & 0 & 1\\ 1 & 0 & 0 & 0 & 0 & 0\\ 0 & 1 & 0 & 0 & 0 & 0\\ 0 & 0 & 1 & 0 & 0 & 0\end{pmatrix}$\\
\addlinespace
\bottomrule
\end{tabular}
\caption{Modular transformations after $\Z{3}$ orbifolding: $\mathrm{SL}(2,\Z{})_T$ modular 
transformations of the $T$-modulus are generated by $\hat{K}_\mathrm{S}$ and $\hat{K}_\mathrm{T}$ 
and extended by the $\CP$-like transformation $\hat{K}_*$, see 
section~\ref{sec:Z3ModularTransformations}. Here and in the following, 
$\omega:=\mathrm{e}^{2\pi\I/3}$}.
\label{tab:ModularTrafoTwistedStrings}
\end{center}
\end{table}

Now, we can use the results of appendix~\ref{sec:BulkVertexOperators} to compute the transformation 
of orbifold-invariant untwisted strings $V(\hat{N})^\mathrm{orb.}$ under the generators of 
unbroken modular transformations
\begin{equation}\label{eq:ModSymmetryGenerators}
\hat{\Sigma} ~\in~ \{\hat{K}_\mathrm{S}, \hat{K}_\mathrm{T}, \hat{K}_* \}\;.
\end{equation}
These modular transformations act naturally on untwisted strings $V(\hat{N})^\mathrm{orb.}$, i.e.\
\begin{equation}
V(\hat{N})^\mathrm{orb.} ~\mapsto~ V(\hat{\Sigma}^{-1}\hat{N})^\mathrm{orb.}\;.
\end{equation}
Hence, modular transformations $\hat{\Sigma}$ in general permute untwisted strings, as indicated by 
the transformations of the classes $V^{(M,N)}$ of untwisted strings given in the third column of 
table~\ref{tab:ModularTrafoTwistedStrings}.

Finally, we compute the transformation properties of twisted strings $(X,\,Y,\,Z)$ and 
$(\bar{X},\,\bar{Y},\,\bar{Z})$. To do so, we examine the OPEs~\cite{Lauer:1989ax,Lauer:1990tm} 
between twisted strings $(X,\,Y,\,Z)$ and $(\bar{X},\,\bar{Y},\,\bar{Z})$ which yield the classes 
$V^{(M,N)}$ of untwisted strings, see eq.~\eqref{eqn:InvertedOPEs} in 
appendix~\ref{sec:Z3VertexOperators}. Since we know the transformations of $V^{(M,N)}$, we can 
infer the transformations of the twisted strings, where possible phases have been fixed using three 
approaches that all lead to the same result: i) By requiring minimality, such that the flavor 
groups computed in section~\ref{sec:FlavorLandscape} are as small as possible; ii) by requiring 
that the resulting flavor groups are identical for equivalent regions in moduli space; and iii) by 
using ref.~\cite{Lerche:1989cs} (ignoring those contributions that originate from other parts of 
the full string vertex operator and the contribution that depends on the $T$-modulus). Doing so, we 
obtain the six-dimensional transformation matrices $\hat{\Sigma}_{\rep{6}}$ of the six twisted 
strings $(X,\,Y,\,Z,\,\bar{X},\,\bar{Y},\,\bar{Z})$ under 
$(\hat{\Sigma},0) \not\in \hat{S}_\mathrm{Narain}$, see 
also~\cite{Lauer:1990tm,Lerche:1989cs,Ferrara:1989qb}. The results are listed in the last column of 
table~\ref{tab:ModularTrafoTwistedStrings}. 

At low energies one can integrate out all massive strings and the effective low-energy theory 
depends only on the massless strings given by the untwisted string $V(\hat{N}=0)^\mathrm{orb.}$ 
and the twisted strings. Under modular transformations with $\hat{K}_{\mathrm{S}}$ and 
$\hat{K}_{\mathrm{T}}$ the untwisted string $V(0)^\mathrm{orb.}$ is invariant, while the twisted 
strings transform with the six-dimensional matrices $\hat{K}_{\mathrm{S}, \rep{6}}$ and 
$\hat{K}_{\mathrm{T}, \rep{6}}$ as given in table~\ref{tab:ModularTrafoTwistedStrings}. It turns 
out that these matrices generate the finite modular group $\mathrm{T}'$, which is the double 
covering group of $A_4 \cong \Gamma_3$. Including the \CP-like transformation $\hat{K}_*$, the 
finite modular group $\mathrm{T}' \cong \mathrm{SL}(2,3)$ is enhanced to $\mathrm{GL}(2,3)$, see 
appendix~\ref{sec:TprimeDecomposition} for further details. In summary, the modular group of 
massless strings on the two-dimensional $\Z{3}$ orbifold is $\mathrm{GL}(2,3)$, a group of order 
48, and it includes the $\CP$-like transformation $\hat{K}_*$.

\subsection[Classification of flavor symmetries]{\boldmath Classification of flavor symmetries\unboldmath}
\label{sec:Z3AllFlavorSymmetries}

Let us now give a complete classification of all outer automorphisms of the $\Z3$ Narain space 
group of type (1.) and (2.b), i.e.\ that leave the moduli invariant, at least at some points in 
moduli space. By doing so, we obtain the unified flavor symmetry of the symmetric $\Z{3}$ orbifold 
in two dimensions. As stated in section~\ref{sec:AutomorphismsNarainSpaceGroup}, the group of outer 
automorphisms can be generated in our case by pure twists $\hat{h}=(\hat\Sigma, 0)\not\in \hat{S}_\mathrm{Narain}$ 
and pure translations $\hat{h}=(\Id, \hat{T})\not\in \hat{S}_\mathrm{Narain}$. Moreover, according 
to section~\ref{sec:ClassesOfTwists}, there are four classes of twist outer automorphisms: symmetric 
rotations, symmetric reflections, asymmetric rotations, and asymmetric reflections. In the following, 
we will discuss the pure translations and these four classes of twists individually.

\begin{figure}[ht]
\centering
\includegraphics[width=0.8\linewidth]{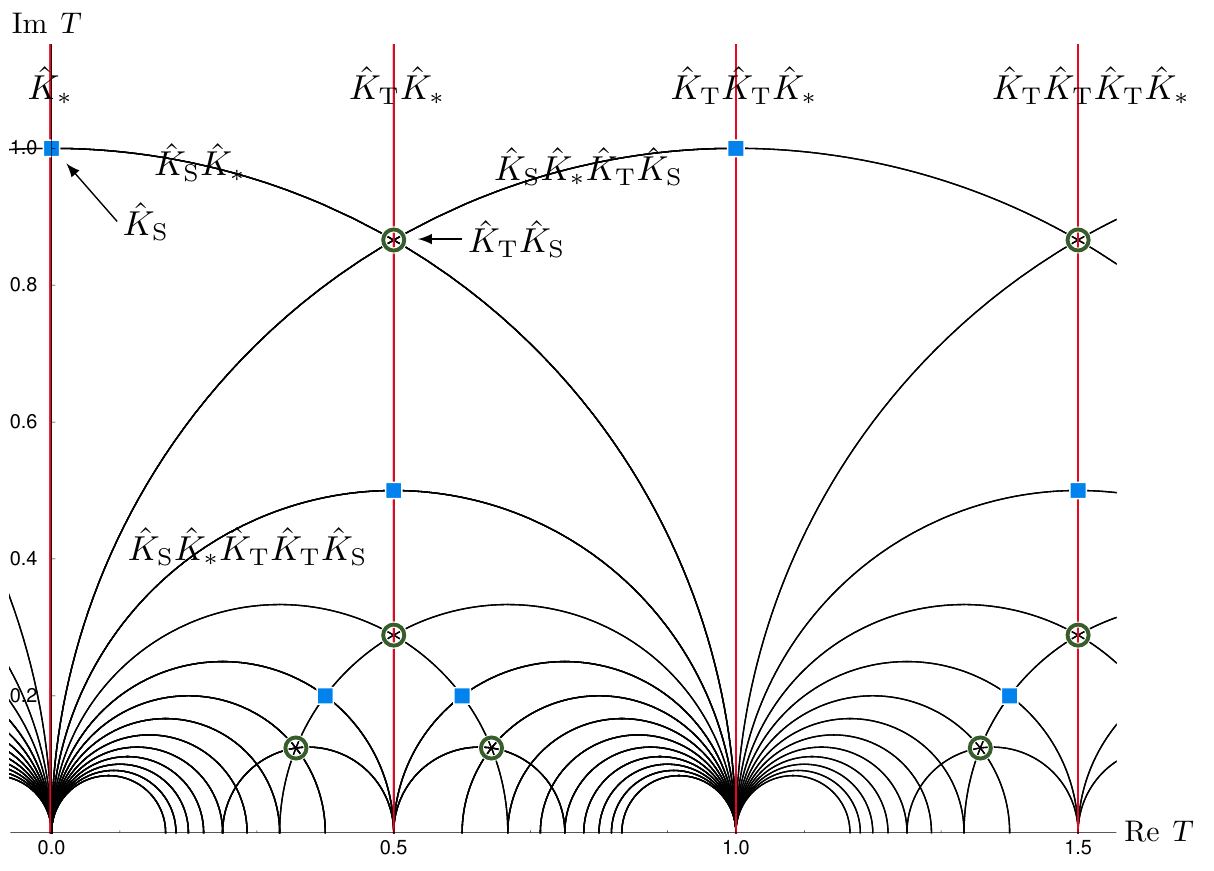}
\caption{\label{fig:modular_space2}
Fixed points and fixed curves in the moduli space of the K\"ahler modulus 
$T=b + \I\, \nicefrac{\sqrt{3}}{2}\,r$ under modular transformations composed out of the generators 
$\hat{K}_\mathrm{S}$, $\hat{K}_\mathrm{T}$, and $\hat{K}_*$. }
\end{figure}

\subsubsection[Pure translations]{\boldmath Pure translations\unboldmath} 
\label{sec:Z3PureTranslations}

The pure translational outer automorphisms $\hat{h}=(\Id, \hat{T})$ with $\hat{T} \not\in \Z{}^4$ 
can be determined by
\begin{equation}
\left(\Id - \hat\Theta\right)\,\hat{T} ~\in~ \Z{}^4\;,
\end{equation}
using eq.~\eqref{eq:ConditionOnOuterAutomorphismOfNarain} with $k=1$. It turns out that there are 
two translations, denoted by $\mathrm{A}$ and $\mathrm{B}$,
\begin{equation}
\mathrm{A} ~=~ (\Id, \hat{T}_1)\;,\;  \mathrm{B} ~=~ (\Id, \hat{T}_2)  \quad\mathrm{with}\quad \hat{T}_1 ~=~ \begin{pmatrix}\frac{1}{3}\\\frac{2}{3}\\0\\0\end{pmatrix} \;,\; \hat{T}_2 ~=~ \begin{pmatrix}0\\0\\\frac{1}{3}\\\frac{1}{3}\\\end{pmatrix}\;,
\end{equation}
which generate all translational outer automorphisms --- at any point in $T$-moduli space. Hence, 
$\mathrm{A}$ and $\mathrm{B}$ are generators of the traditional flavor symmetry as defined in 
section~\ref{sec:ModularVSFlavor}.

The translation $\mathrm{A}$ shifts the winding number, while $\mathrm{B}$ shifts the KK number. 
Using eq.~\eqref{eqn:CoordinatesYAndTildeY} these shifts can be translated to shifts of the 
geometrical coordinates $y$ and their T-duals $\tilde{y}$: the translation $\mathrm{A}$ shifts $y$ 
(and simultaneously $\tilde{y}$ if the $B$-field is nontrivial), while $\mathrm{B}$ shifts only the 
T-dual coordinate $\tilde{y}$ but leaves the geometrical coordinate $y$ inert. Consequently, the 
translation $\mathrm{B}$ can not be obtained in a purely geometrical approach but only in the 
Narain construction.

\subsubsection[Symmetric rotations]{\boldmath Symmetric rotations\unboldmath} 
\label{sec:Z3SymmetricRotation}

We start with the most general left-right-symmetric rotation in $(2,2)$ dimensions 
$S_\mathrm{rot.}(\alpha)$, where the rotation angle $\alpha$ is constrained crystallographically 
according to eq.~\eqref{eq:OrdersOfRotationIn4D}. By changing the basis to the Narain lattice basis, 
\begin{equation}
\hat{S}_\mathrm{rot.}\!\left(\alpha\right) ~=~ E^{-1}\, S_\mathrm{rot.}(\alpha)\,E ~\stackrel{!}{\in}~ \mathrm{GL}(4,\Z{})\;,
\end{equation}
one finds that left-right-symmetric rotations in $(2,2)$ dimensions are automorphisms of the 
$\Z{3}$ Narain space group only for $\alpha = \nicefrac{2\pi\,\ell}{6}$ and $\ell\in\{0,\ldots,5\}$, 
independent of the value of the $T$-modulus. Hence, one confirms that the order 6 transformation
\begin{equation}\label{eq:SymZ6Rotations}
\hat{S}_\mathrm{rot.}\!\left(\nicefrac{2\pi}{6}\right) ~=~ \begin{pmatrix}1&-1&0&0\\1&0&0&0\\0&0&0&-1\\0&0&1&1\end{pmatrix} ~=~ \left(\hat{C}_\mathrm{S} \hat{C}_\mathrm{T}\right)^5\quad\mathrm{with}\quad \left(\hat{S}_\mathrm{rot.}\!\left(\nicefrac{2\pi}{6}\right)\right)^6 ~=~ \Id\;,
\end{equation}
generates all symmetric rotations. Since 
$\left(\hat{S}_\mathrm{rot.}\!\left(\nicefrac{2\pi}{6}\right)\right)^2 = \hat\Theta$ is the 
orbifold twist $\hat\Theta$, see eq.~\eqref{eq:Z3TwistLatticeBasis}, this $\Z{6}$ can be written as 
$\Z{3}\times\Z{2}$ generated by the inner automorphism $\hat\Theta$ and the outer automorphism 
\begin{equation}\label{eq:SymZ2Rotation}
\mathrm{C}~:=~ (\hat{S}_\mathrm{rot.}(\pi), 0) \quad\mathrm{where}\quad \hat{S}_\mathrm{rot.}(\pi) ~=~ \left(\hat{S}_\mathrm{rot.}\!\left(\nicefrac{2\pi}{6}\right)\right)^3 ~=~ -\Id ~=~ \left(\hat{K}_\mathrm{S}\right)^2 \;.
\end{equation}
We denote this rotational outer automorphism $(\hat{S}_\mathrm{rot.}(\pi), 0)$ of the $\Z{3}$ 
Narain space group $\hat{S}_\mathrm{Narain}$ by $\mathrm{C}$.

In summary, the symmetric rotation $\mathrm{C} = (\hat{S}_\mathrm{rot.}(\pi), 0)$ is unbroken at 
any point in $T$-moduli space. Hence, $\mathrm{C}$ is an outer automorphism of type (1.) and 
belongs, together with $\mathrm{A}$ and $\mathrm{B}$, to the traditional flavor symmetry as defined 
in section~\ref{sec:ModularVSFlavor}.

\subsubsection[Symmetric reflections]{\boldmath Symmetric reflections\unboldmath} 
\label{sec:Z3SymmetricReflection}

Next, we discuss left-right-symmetric reflections $S_\mathrm{refl.}(\alpha)$ in $(2,2)$ dimensions. 
Again, we change the basis to the Narain lattice basis, 
\begin{equation}
\hat{S}_\mathrm{refl.}\!\left(\alpha\right) ~=~ E^{-1}\, S_\mathrm{refl.}(\alpha)\,E ~\stackrel{!}{\in}~ \mathrm{GL}(4,\Z{})\;.
\end{equation}
Then, one finds that the general left-right-symmetric reflection 
$\hat{S}_\mathrm{refl.}\!\left(\alpha\right)$ is an integer matrix only for 
$\alpha = \nicefrac{2\pi\,\ell}{6}$ with $\ell\in\{0,\ldots,5\}$, for all choices of the radius $r$ 
but only for quantized values of the $B$-field $b=\nicefrac{n_B}{2}$ with $n_B\in\Z{}$. Thus, the 
symmetric reflection $\hat{S}_\mathrm{refl.}\!\left(\alpha\right)$ depends additionally on 
$n_B\in\Z{}$ and we write $\hat{S}_\mathrm{refl.}^{(n_B)}\!\left(\alpha\right)$. One confirms that 
the transformations
\begin{equation}
\hat{S}_\mathrm{refl.}^{(n_B)}\!\left(\alpha\right) ~=~ \left(\hat{S}_\mathrm{rot.}(\pi)\right)^v\, \left(\Hat\Theta\right)^w\, \hat{S}_\mathrm{refl.}^{(n_B)}\!\left(\nicefrac{2\pi}{6}\right)
\end{equation}
describe all symmetric reflections for $v\in\{0,1\}$ and $w\in\{0,1,2\}$ such that the reflection 
axis has an angle $\alpha=\nicefrac{2\pi}{6}(1+3v+2w)$. Consequently, we can choose 
$\hat{S}_\mathrm{refl.}^{(n_B)}\!\left(\nicefrac{2\pi}{6}\right)$ as the sole generator of the 
symmetric reflections, and define
\begin{equation}
\hat{S}_\mathrm{refl.}\!\left(n_B\right) ~:=~ \hat{S}_\mathrm{refl.}^{(n_B)}\!\left(\nicefrac{2\pi}{6}\right) ~=~ \begin{pmatrix}1&0&0&0\\1&-1&0&0\\-n_B&n_B&1&1\\n_B&0&0&-1\end{pmatrix}\;.
\end{equation}
Using section~\ref{sec:ModularSymmetry}, this outer automorphism can be decomposed into the 
generators of the modular group as
\begin{equation}
\hat{S}_\mathrm{refl.}\!\left(n_B\right) ~=~ \left(\hat{K}_\mathrm{T}\right)^{n_B} \hat{K}_* \quad\mathrm{for}\quad n_B ~\in~\Z{}\;.
\end{equation}
This reflective outer automorphism $(\hat{S}_\mathrm{refl.}\!\left(n_B\right), 0)$ of 
$\hat{S}_\mathrm{Narain}$ is denoted by $\mathrm{D}(n_B)$ for $n_B \in\Z{}$.

In summary, the symmetric reflection 
$\mathrm{D}(n_B) = (\hat{S}_\mathrm{refl.}\!\left(n_B\right), 0)$ is unbroken for an arbitrary 
radius $r$ in $T$-moduli space, but only for $b=\nicefrac{n_B}{2}$ with $n_B\in\Z{}$, see 
figure~\ref{fig:modular_space2}. Hence, $\mathrm{D}(n_B)$ is an outer automorphism of type (2.b) 
and belongs to the unified flavor symmetry as defined in section~\ref{sec:ModularVSFlavor}.

\subsubsection[Asymmetric rotations]{\boldmath Asymmetric rotations\unboldmath} 
\label{sec:Z3AsymmetricRotation}

Left-right asymmetric rotations
\begin{equation}
\hat{A}_\mathrm{rot.}\!\left(\alpha_\mathrm{R},\alpha_\mathrm{L}\right) ~=~ E^{-1}\, A_\mathrm{rot.}(\alpha_\mathrm{R},\alpha_\mathrm{L})\,E ~\stackrel{!}{\in}~ \mathrm{GL}(4,\Z{})
\end{equation}
are automorphisms of the $\Z{3}$ Narain space group for two classes of values of the $T$-modulus. 
Firstly, at all points $T_1=b$, $T_2=\nicefrac{1}{c}$ that fulfill
\begin{equation}\label{eq:LRAsymmetricClass1}
c,b \cdot c~\in~\Z{} \quad\mathrm{and}\quad b^2\,c+\frac{1}{c} ~\in~ \Z{}\;,
\end{equation}
the left-right asymmetric rotations are generated by the transformation
\begin{equation}
\hat{A}_\mathrm{rot.}\!\left(\nicefrac{2\pi\,7}{12},\nicefrac{2\pi}{12}\right) ~=~ c\,\begin{pmatrix}0&b&1&0\\-b&b&1&1\\b^2+\frac{1}{c^2}&-b^2-\frac{1}{c^2}&-b&-b\\0&b^2+\frac{1}{c^2}&b&0\end{pmatrix}\;.
\end{equation}
Secondly, at all points $T_1=b$, $T_2=\nicefrac{\sqrt{3}\,r}{2}$ that fulfill
\begin{equation}\label{eq:LRAsymmetricClass2}
\frac{1}{r},\frac{|T|^2}{r} ~\in~ \Z{} \quad\mathrm{and}\quad \frac{2\,b}{r} ~\in~ \Z{}^{\mathrm{odd}}\;,
\end{equation}
the left-right asymmetric rotations are generated by the symmetric $\Z{6}$ rotation 
$\hat{S}_\mathrm{rot.}\!(\nicefrac{2\pi}{6})$ and the transformation
\begin{equation}
\hat{A}_\mathrm{rot.}\!\left(\nicefrac{2\pi}{3},0\right) ~=~ \begin{pmatrix}-\frac{b}{r}+\frac{1}{2}&\frac{b}{r}-\frac{1}{2}&\frac{1}{r}&\frac{1}{r}\\[2pt]-\frac{b}{r}+\frac{1}{2}&0&0&\frac{1}{r}\\[2pt]\frac{b^2}{r}+\frac{3r}{4}&0&0&-\frac{b}{r}-\frac{1}{2}\\[2pt]-\frac{b^2}{r}-\frac{3r}{4}&\frac{b^2}{r}+\frac{3r}{4}&\frac{b}{r}+\frac{1}{2}&\frac{b}{r}+\frac{1}{2}\end{pmatrix}\;.
\end{equation}
Together with the already discussed outer automorphisms, the two asymmetric rotations above 
generate all possible asymmetric rotations. For example, at first sight 
$\hat{A}_\mathrm{rot.}\!\left(0,\nicefrac{2\pi}{3}\right)$ is an additional left-right asymmetric 
transformation. However, this transformation is not independent since
\begin{equation}
\hat{A}_\mathrm{rot.}\!\left(0,\nicefrac{2\pi}{3}\right) ~=~ \left(\hat{A}_\mathrm{rot.}\!\left(\nicefrac{2\pi}{3}, 0\right)\right)^2 \, \left(\hat{S}_\mathrm{rot.}\!\left(\nicefrac{2\pi}{6}\right)\right)^2\;.
\end{equation}
The points of the first class eq.~\eqref{eq:LRAsymmetricClass1} correspond to the blue squares in 
figure~\ref{fig:modular_space2}, while the second class eq.~\eqref{eq:LRAsymmetricClass2} is 
located at the green curls. 

Let us give one example per class: one solution of eq.~\eqref{eq:LRAsymmetricClass1} is given by 
$b=0$ and $r=\nicefrac{2}{\sqrt{3}}$ (i.e.\ $c=1$) yielding 
$\hat{A}_\mathrm{rot.}\!\left(\nicefrac{2\pi\,7}{12},\nicefrac{2\pi}{12}\right) = \hat{K}_\mathrm{S}\,\hat{\Theta}$, 
while one solution of eq.~\eqref{eq:LRAsymmetricClass2} is given by $b=\nicefrac{1}{2}$ and $r=1$ 
resulting in $\hat{A}_\mathrm{rot.}\!\left(\nicefrac{2\pi}{3},0\right) = \hat{K}_\mathrm{T}\,\hat{K}_\mathrm{S}\,\hat{S}_\mathrm{rot.}\!\left(\nicefrac{2\pi}{6}\right)$. 
Since $\hat\Theta$ in the first class is an inner automorphism and 
$\hat{S}_\mathrm{rot.}\!\left(\nicefrac{2\pi}{6}\right)$ in the second class is part of the 
traditional flavor symmetry, see section~\ref{sec:Z3SymmetricRotation}, the corresponding 
enhancements are entirely generated by $\hat{K}_\mathrm{S}$ and 
$\hat{K}_\mathrm{T}\,\hat{K}_\mathrm{S}$, respectively.

In summary, there are two independent asymmetric rotations 
$(\hat{A}_\mathrm{rot.}\!\left(\nicefrac{2\pi\,7}{12},\nicefrac{2\pi}{12}\right), 0)$ 
and $(\hat{A}_\mathrm{rot.}\!\left(0,\nicefrac{2\pi}{3}\right), 0)$. 
Both are unbroken only at special points in $T$-moduli space, given by 
eq.~\eqref{eq:LRAsymmetricClass1} and eq.~\eqref{eq:LRAsymmetricClass2}. Hence, these asymmetric 
rotations are outer automorphisms of type (2.b) and, therefore, contribute to the unified flavor 
symmetries as defined in section~\ref{sec:ModularVSFlavor}.

\subsubsection[Asymmetric reflections]{\boldmath Asymmetric reflections\unboldmath} 
\label{sec:Z3AsymmetricReflection}

Finally, left-right-asymmetric reflections
\begin{equation}
\hat{A}_\mathrm{refl.}\!\left(\alpha_\mathrm{R},\alpha_\mathrm{L}\right) ~=~ E^{-1}\, A_\mathrm{refl.}(\alpha_\mathrm{R},\alpha_\mathrm{L})\,E ~\stackrel{!}{\in}~ \mathrm{GL}(4,\Z{}) 
\end{equation}
are automorphisms of the $\Z{3}$ Narain space group only for special values of the $T$-modulus and 
special angles $\alpha_\mathrm{R} \neq \alpha_\mathrm{L}$ of the reflection axis. One can confirm that 
the transformations
\begin{equation}\label{eq:AsymmetricReflections}
\hat{A}_\mathrm{refl.}\!\left(\alpha_\mathrm{R},\alpha_\mathrm{L}\right) ~=~ \begin{pmatrix}-w&0&0&v\\-w&w&v&v\\-\frac{1-w^2}{v}&\frac{1-w^2}{v}&-w&-w\\\frac{1-w^2}{v}&0&0&w\end{pmatrix}
\end{equation}
describe all asymmetric reflections up to $\Z{6}$ rotations given in eq.~\eqref{eq:SymZ6Rotations} for
\begin{equation}\label{eq:ConditionsForAsymmetricReflections}
v,w ~\in~\Z{} \quad\mathrm{ and }\quad \frac{1-w^2}{v} ~\in~ \Z{}\;.
\end{equation}
Furthermore, the $T$-modulus is constrained to live on a circle of radius $\nicefrac{1}{|v|}$ with 
center at $(\nicefrac{w}{v},0)$,
\begin{equation}\label{eq:Tcircles}
\left(T_1-\frac{w}{v}\right)^2 + \left(T_2\right)^2 ~=~ \frac{1}{v^2}\;.
\end{equation}

For example, a solution of eq.~\eqref{eq:ConditionsForAsymmetricReflections} is given by $v=\pm 1$ 
and $w=0$ yielding a circle of radius $1$, centered at $(0,0)$. In this case 
eq.~\eqref{eq:AsymmetricReflections} is given by $\hat{A}_\mathrm{refl.1} := (\hat{K}_\mathrm{S})^3 \hat{K}_*$ 
for $v=1$ and $\hat{A}_\mathrm{refl.1'} := \hat{K}_\mathrm{S} \hat{K}_*$ for $v=-1$.

In summary, the asymmetric reflection $(\hat{A}_\mathrm{refl.}\!\left(\alpha_\mathrm{R},\alpha_\mathrm{L}\right), 0)$ 
is unbroken only on the circles eq.~\eqref{eq:Tcircles} in $T$-moduli space, see 
figure~\ref{fig:modular_space2}. Hence, 
$(\hat{A}_\mathrm{refl.}\!\left(\alpha_\mathrm{R},\alpha_\mathrm{L}\right), 0)$ belongs to the 
unified flavor symmetry as defined in section~\ref{sec:ModularVSFlavor}.

\section[Unified flavor symmetries of the Z3 orbifold]{\boldmath Unified flavor symmetries of the $\Z{3}$ orbifold\unboldmath}
\label{sec:FlavorLandscape}

After having classified the generators of the (traditional and unified) flavor symmetries for the 
$\Z{3}$ orbifold on the level of outer automorphisms of the corresponding Narain space group in 
section~\ref{sec:Z3AllFlavorSymmetries}, we determine the resulting, moduli-dependent flavor 
symmetries in this section. To do so, it is necessary to identify the transformation properties of 
twisted strings in order to get faithful representations of the resulting flavor groups.

\subsection[Traditional flavor symmetry at a generic point in <T>: Delta(54)]{\boldmath Traditional flavor symmetry at a generic point in $\langle T\rangle$: $\Delta(54)$\unboldmath} 
\label{sec:GenericPoint}

The outer automorphisms $\hat{h}$ of the $(2,2)$-dimensional $\Z{3}$ Narain space group of type 
(1.), i.e.\ the ones that are unbroken at a generic point $\langle T\rangle$ in $T$-moduli space, 
can be generated by two translations $\mathrm{A}$ and $\mathrm{B}$, defined in 
section~\ref{sec:Z3PureTranslations}, and one symmetric $\Z{2}$ rotation $\mathrm{C}$, defined in 
section~\ref{sec:Z3SymmetricRotation}. In order to identify the actual symmetry group of the 
traditional flavor symmetry, we determine the transformation properties of untwisted and twisted 
strings under these actions in the following.

First, we consider an orbifold-invariant untwisted string $V(\hat{N})^\mathrm{orb.}$ with winding and 
KK charges $N,M \in \{0,1,2\}$ such that $\hat{N} \in \Gamma_{MN} \subset \Z{}^4$. Then, using 
eq.~\eqref{eqn:TrafoOfBulkVertexOperator} from appendix~\ref{sec:BulkVertexOperators} we obtain the 
following transformation properties with respect to the translational generators of the outer 
automorphisms $\mathrm{A}$ and $\mathrm{B}$ and with respect to the rotation $\mathrm{C}$:
\begin{subequations}\label{eqn:BulkDelta54Trafo}
\begin{eqnarray}
V(\hat{N})^\mathrm{orb.} & \stackrel{\mathrm{A}}{\longmapsto} & \omega^{2M}\, V(\hat{N})^\mathrm{orb.}\;,\\
V(\hat{N})^\mathrm{orb.} & \stackrel{\mathrm{B}}{\longmapsto} & \omega^{N}\,  V(\hat{N})^\mathrm{orb.}\;,\\
V(\hat{N})^\mathrm{orb.} & \stackrel{\mathrm{C}}{\longmapsto} & V(-\hat{N})^\mathrm{orb.}\;,
\end{eqnarray}
\end{subequations}
where $\omega = \exp\nicefrac{2\pi\I}{3}$. Consequently, an untwisted string $V(\hat{N})^\mathrm{orb.}$ 
with $\hat{N} \in \Gamma_{MN}$ transforms with phases $\omega^{2M}$ and $\omega^{N}$ under the 
translations $\mathrm{A}$ and $\mathrm{B}$, respectively, and (for $\hat{N}\neq 0$) gets 
interchanged with $V(-\hat{N})^\mathrm{orb.}$ under $\mathrm{C}$.

\begin{table}[t!]
\begin{center}
\begin{tabular}{c|c|c|ccc}
\toprule
irrep       & untw. vertex operator & $(M,N)$ & $\mathrm{A}$  & $\mathrm{B}$  & $\mathrm{C}$  \\ 
\midrule
\addlinespace
$\rep{1}_0$ & $V(\hat{N})^\mathrm{orb.} + V(-\hat{N})^\mathrm{orb.}$ & $(0,0)$ & $1$  & $1$  & $1$  \\
%\addlinespace
$\rep{1}'$  & $V(\hat{N})^\mathrm{orb.} - V(-\hat{N})^\mathrm{orb.}$ & $(0,0)$ & $1$  & $1$  & $-1$ \\
\addlinespace
\midrule[0.15mm]
\addlinespace$\rep{2}_1$ & $\begin{pmatrix} V(\hat{N})^\mathrm{orb.}\\ V(-\hat{N})^\mathrm{orb.}\end{pmatrix}$ & $\begin{array}{c}(0,2)\\(0,1)\end{array}$ & $\begin{pmatrix} 1&0\\0&1\end{pmatrix}$ & $\begin{pmatrix} \omega^2&0\\0&\omega\end{pmatrix}$ & $\begin{pmatrix} 0&1\\1&0\end{pmatrix}$ \\[11pt]
%\addlinespace
$\rep{2}_2$ & $\begin{pmatrix} V(\hat{N})^\mathrm{orb.}\\ V(-\hat{N})^\mathrm{orb.}\end{pmatrix}$ & $\begin{array}{c}(1,0)\\(2,0)\end{array}$ & $\begin{pmatrix} \omega^2&0\\0&\omega\end{pmatrix}$ & $\begin{pmatrix} 1&0\\0&1\end{pmatrix}$ & $\begin{pmatrix} 0&1\\1&0\end{pmatrix}$ \\[11pt]
%\addlinespace
$\rep{2}_3$ & $\begin{pmatrix} V(\hat{N})^\mathrm{orb.}\\ V(-\hat{N})^\mathrm{orb.}\end{pmatrix}$ & $\begin{array}{c}(1,2)\\(2,1)\end{array}$ & $\begin{pmatrix} \omega^2&0\\0&\omega\end{pmatrix}$ & $\begin{pmatrix} \omega^2&0\\0&\omega\end{pmatrix}$ & $\begin{pmatrix} 0&1\\1&0\end{pmatrix}$ \\[11pt]
%\addlinespace
$\rep{2}_4$ & $\begin{pmatrix} V(\hat{N})^\mathrm{orb.}\\ V(-\hat{N})^\mathrm{orb.}\end{pmatrix}$ & $\begin{array}{c}(1,1)\\(2,2)\end{array}$ & $\begin{pmatrix} \omega^2&0\\0&\omega\end{pmatrix}$ & $\begin{pmatrix} \omega&0\\0&\omega^2\end{pmatrix}$ & $\begin{pmatrix} 0&1\\1&0\end{pmatrix}$ \\
%\addlinespace
\midrule
irrep       & twisted vertex operator  & $N$ & $\mathrm{A}$  & $\mathrm{B}$  & $\mathrm{C}$  \\ 
\midrule
\addlinespace
$\rep{3}_2$ & $\begin{pmatrix} X\\ Y\\ Z\end{pmatrix}$ & $\begin{array}{c}0\\1\\2\end{array}$ & $\begin{pmatrix} 0&1&0\\0&0&1\\1&0&0\end{pmatrix}$ & $\begin{pmatrix} 1&0&0\\0&\omega&0\\0&0&\omega^2\end{pmatrix}$ & $\begin{pmatrix} -1&0&0\\0&0&-1\\0&-1&0\end{pmatrix}$ \\
%\addlinespace
$\crep{3}_2$& $\begin{pmatrix} \bar{X}\\ \bar{Y}\\ \bar{Z}\end{pmatrix}$ & $\begin{array}{c}0\\2\\1\end{array}$ & $\begin{pmatrix} 0&1&0\\0&0&1\\1&0&0\end{pmatrix}$ & $\begin{pmatrix} 1&0&0\\0&\omega^2&0\\0&0&\omega\end{pmatrix}$ & $\begin{pmatrix} -1&0&0\\0&0&-1\\0&-1&0\end{pmatrix}$ \\
\addlinespace
\bottomrule
\end{tabular}
\caption{$\Delta(54)$ transformation properties of $\Z{3}$ orbifold-invariant untwisted strings 
$V(\hat{N})^\mathrm{orb.}$ and twisted strings $(X,\,Y,\,Z)$ and $(\bar{X},\,\bar{Y},\,\bar{Z})$ 
from the first and second twisted sector, respectively. Note that $\hat{N} \in \Gamma^{(M,N)}$ 
and the case $\hat{N} = 0$ is excluded for $\rep{1}'$.}
\label{tab:Delta54Bulk}
\end{center}
\end{table}

Next, we compute the transformation properties of twisted strings 
$(X,\,Y,\,Z,\,\bar{X},\,\bar{Y},\,\bar{Z})$ under $\mathrm{A}$, $\mathrm{B}$, and $\mathrm{C}$. For 
the symmetric rotation $\mathrm{C}$, we can translate eq.~\eqref{eq:SymZ2Rotation} into the 
six-dimensional representation 
$\hat{S}_\mathrm{rot.}(\pi) =(\hat{K}_\mathrm{S})^2 \mapsto (\hat{K}_{\mathrm{S},\rep{6}})^2$ using 
table~\ref{tab:ModularTrafoTwistedStrings} and obtain 
\begin{equation}
\begin{pmatrix}X\\Y\\Z\\\bar{X}\\\bar{Y}\\\bar{Z}\end{pmatrix} ~\stackrel{~\mathrm{C}~}{\longmapsto}~ \begin{pmatrix}-1&0&0&0&0&0\\0&0&-1&0&0&0\\0&-1&0&0&0&0\\0&0&0&-1&0&0\\0&0&0&0&0&-1\\0&0&0&0&-1&0\end{pmatrix}\,\begin{pmatrix}X\\Y\\Z\\\bar{X}\\\bar{Y}\\\bar{Z}\end{pmatrix}\;.
\end{equation}
On the other hand, for the translations $\mathrm{A}$ and $\mathrm{B}$ we use the 
OPEs~\eqref{eqn:InvertedOPEs} from appendix~\ref{sec:Z3VertexOperators} to identify the 
transformation properties of twisted strings from the corresponding ones of untwisted strings. 
Hence, the strategy is exactly complementary to the one used in ref.~\cite{Nilles:2018wex} where 
the transformation of doublets was extracted from the assumed transformation properties of the 
triplets. As a result, the six-dimensional representation of the outer automorphisms $\mathrm{A}$, 
$\mathrm{B}$, and $\mathrm{C}$ generate the traditional flavor symmetry $\Delta(54)$. The results 
are summarized in table~\ref{tab:Delta54Bulk}. 

Note that $\mathrm{C}$ corresponds to a rotation in extra dimensions, hence, one is tempted to 
interpret $\mathrm{C}$ as an $R$-symmetry of $\mathcal{N}=1$ supersymmetry. Consequently, 
$\Delta(54)$ gets promoted to the first example of a non-Abelian discrete 
$R$-symmetry~\cite{Chen:2013dpa} from strings (where the Grassmann variable of $\mathcal{N}=1$ 
superspace transforms in a nontrivial one-dimensional representation). A detailed discussion of the 
transformation of world-sheet fermions under the rotation $\mathrm{C}$ is needed to settle this 
question.

Moreover, since $\mathrm{A}$ and $\mathrm{B}$ are defined as translational outer automorphisms they 
seem to commute and generate $\Z{3}\times\Z{3}$ on first sight. This would have been true if they 
had affected only the untwisted strings, cf.\ the simultaneously diagonal operators for doublets in 
table~\ref{tab:Delta54Bulk}. However, by analyzing the action of $\mathrm{A}$ and $\mathrm{B}$ on 
twisted strings, c.f.\ table~\ref{tab:Delta54Bulk}, we realize that $\mathrm{A}$ and $\mathrm{B}$ 
do not commute. Interestingly, $\mathrm{A^2B^2AB}$ and $\mathrm{B}$ \textit{do} commute and it is 
precisely them which give rise to the well-known $\Z{3}\times\Z{3}$ point and space group selection 
rules~\cite{Hamidi:1986vh,Ramos-Sanchez:2018edc}.

\subsection[Unified flavor symmetry at b=integer/2: SG(108,17)]{\boldmath Unified flavor symmetry at $b=\frac{\mathrm{integer}}{2}$: $\mathrm{SG}(108,17)$\unboldmath}
\label{sec:HalfIntB}

Let us now discuss the possible enhancements of the traditional flavor symmetry by modular 
transformations, which together give rise to the unified flavor symmetry. For quantized values of 
the $B$-field $b=\frac{n_B}{2}$ with $n_B\in\Z{}$ and generic radii $r$ the traditional flavor 
symmetry $\Delta(54)$ gets enhanced by a left-right-symmetric reflection 
$\mathrm{D}(n_B)=(\hat{S}_\mathrm{refl.}(n_B),0)$, where 
$\hat{S}_\mathrm{refl.}(n_B) = (\hat{K}_\mathrm{T})^{n_B} \hat{K}_*$, as described in 
section~\ref{sec:Z3SymmetricReflection}. 

In order to determine the resulting unified flavor symmetry, we use 
table~\ref{tab:ModularTrafoTwistedStrings} to construct the six-dimensional (faithful) 
representation $(\hat{K}_{\mathrm{T},\rep{6}})^{n_B} \hat{K}_{*,\rep{6}}$ of 
$\hat{S}_\mathrm{refl.}\!(n_B)$ that acts on the six twisted strings, i.e.
\begin{equation}
\begin{pmatrix}X\\Y\\Z\\\bar{X}\\\bar{Y}\\\bar{Z}\end{pmatrix} ~\stackrel{~\hat{S}_\mathrm{refl.}(n_B)_{\phantom{I_I}}}{\longmapsto}~ \begin{pmatrix}0&0&0&\omega^{2n_B}&0&0\\0&0&0&0&1&0\\0&0&0&0&0&1\\\omega^{n_B}&0&0&0&0&0\\0&1&0&0&0&0\\0&0&1&0&0&0\end{pmatrix}\,\begin{pmatrix}X\\Y\\Z\\\bar{X}\\\bar{Y}\\\bar{Z}\end{pmatrix}\;.
\end{equation}
Note that this six-dimensional representation of $\hat{S}_\mathrm{refl.}\!(n_B)$ has a 
mod-3-periodicity in $n_B$, i.e.\ $\hat{S}_{\mathrm{refl.}, \rep{6}}(n_B+3) = \hat{S}_{\mathrm{refl.}, \rep{6}}(n_B)$. 
Hence, there are three different $\Z{2}$ enhancements of $\Delta(54)$. In all three cases, one 
finds that the enhanced symmetry is $\mathrm{SG}(108,17)$, see figure~\ref{fig:modular_space1}. For 
example, this can done by using GAP \cite{GAP4}. We note that $\mathrm{SG}(108,17)$ is a group of 
\CP-type IIA --- in contrast to $\Delta(54)$ which is a \CP-type I 
group~\cite{Chen:2014tpa,Trautner:2016ezn}.

Consequently, we obtain a coherent picture of spontaneous \CP-breaking: using the results from 
section~\ref{sec:ModularSymmetry} the $T$-modulus transforms under $\hat{S}_\mathrm{refl}(n_B)$ as 
$T \longmapsto n_B - \overline{T}$. Hence, the VEV $\langle T\rangle$ is invariant for quantized 
$B$-field $b=\frac{n_B}{2}$, i.e.
\begin{equation}
\langle T\rangle ~\mapsto~ n_B - \overline{\langle T\rangle} ~=~ \langle T\rangle \quad\mathrm{for}\quad \langle T\rangle ~=~ \frac{n_B}{2} + \I\, \frac{\sqrt{3}}{2}\,\langle r\rangle\;.
\end{equation}
Here, $n_B\in\Z{}$ specifies lines in $T$-moduli space, see figure~\ref{fig:modular_space2}, where 
the traditional flavor symmetry $\Delta(54)$ is enhanced to $\mathrm{SG}(108,17)$ and the \CP-like 
transformation $\hat{S}_\mathrm{refl}(n_B)$ is unbroken. Furthermore, moving $\langle T\rangle$ 
away from the symmetry-enhanced lines,
\begin{equation}
\langle T\rangle~=~\frac{n_B}{2}+\I\,\frac{\sqrt{3}}{2}\,\langle r\rangle\qquad\mathrm{to}\qquad\langle T'\rangle~=~\langle T\rangle+\delta T\qquad\mathrm{where}\qquad\mathrm{Re}(\delta T)\neq 0\;,
\end{equation}
leads to spontaneous symmetry breaking of the flavor group $\mathrm{SG}(108,17)$ to $\Delta(54)$, 
i.e.\ the \CP-like transformation $\hat{S}_\mathrm{refl}(n_B)$ gets broken 
spontaneously~\cite{Baur:2019kwi}. This is another example of spontaneous breaking of \CP where a 
\CP-type IIA group gets broken to a group of \CP-type I, implying that \CP will be violated 
by quantized geometrical phases \cite{Nilles:2018wex}.

\subsection[Unified flavor symmetry at |T|^2 = 1: SG(108,17)]{\boldmath Unified flavor symmetry at $|T|^2 = 1$: $\mathrm{SG}(108,17)$\unboldmath}
\label{sec:TsquaredOne}

A similar picture emerges on special circles in $T$-moduli space. For example, let us discuss the 
semi-circle $|T|^2 = 1$ with $T_2 > 0$ in figure~\ref{fig:modular_space2}. There, the traditional 
flavor symmetry $\Delta(54)$ gets enhanced by two left-right-asymmetric reflections 
$\hat{A}_\mathrm{refl.1} := (\hat{K}_\mathrm{S})^3\, \hat{K}_*$ and 
$\hat{A}_\mathrm{refl.1'} := \hat{K}_\mathrm{S} \hat{K}_*$ corresponding to $v=\pm 1$ and $w=0$, 
respectively, see section~\ref{sec:Z3AsymmetricReflection}. Since 
$\hat{A}_\mathrm{refl.1'} = \hat{S}_\mathrm{rot.}(\pi)\, \hat{A}_\mathrm{refl.1}$ and 
$\hat{S}_\mathrm{rot.}(\pi)$ is contained in $\Delta(54)$, these reflections are not independent. 

In order to identify the unified flavor symmetry, we use 
table~\ref{tab:ModularTrafoTwistedStrings} to construct the six-dimensional (faithful) 
representation of
\begin{equation}
\hat{A}_\mathrm{refl.1} ~=~ \left(\hat{K}_{\mathrm{S}}\right)^3 \hat{K}_{*} ~\mapsto~ \left(\hat{K}_{\mathrm{S},\rep{6}}\right)^3 \hat{K}_{*,\rep{6}}\;.
\end{equation}
Again using GAP~\cite{GAP4}, we find that this $\Z{2}$ outer automorphism enhances the traditional 
flavor group $\Delta(54)$ to $\mathrm{SG}(108,17)$, see figure~\ref{fig:modular_space1}. Hence, 
similar to the \CP-like transformations that are conserved on straight lines in $T$-moduli space, 
for example $\hat{S}_\mathrm{refl}(0)$ at $b=0$, there are three \CP-like transformations that are 
conserved on the respective circles. These three transformations correspond to three more $\Z{2}$ 
outer automorphisms of $\Delta(54)$ which are conjugate to the three previously identified $\Z{2}$ 
in full $\mathrm{S}_4$ group of outer automorphisms of $\Delta(54)$. Consequently, the unified 
flavor symmetry is $\mathrm{SG}(108,17)$ on all of these lines and circles. However, the 
respective groups $\mathrm{SG}(108,17)$ are not identical, but conjugate to each other.

\subsection[Unified flavor symmetry at b=0 and r=2/sqrt(3): SG(216,87)]{\boldmath Unified flavor symmetry at $b=0$ and $r=\nicefrac{2}{\sqrt{3}}$: $\mathrm{SG}(216,87)$\unboldmath}
\label{sec:SG216}

Let us now consider the point $(b,r)=(0,\nicefrac{2}{\sqrt{3}})$ in figure~\ref{fig:modular_space2}. 
At this point, we identify the following type (1.) and type (2.b) outer automorphisms of the Narain 
space group:
\begin{enumerate}
\item $\Delta(54)$ is the traditional flavor symmetry at a generic point $(b,r)$ in $T$-moduli 
space, see section~\ref{sec:GenericPoint}.

\item On the line at $b=0$ we find a left-right-symmetric reflection 
$\hat{S}_\mathrm{refl.}(0) = \hat{K}_*$, see section~\ref{sec:Z3SymmetricReflection} for $n_B=0$.

\item From the point $(b,r)=(0,\nicefrac{2}{\sqrt{3}})$ we get a $\Z{4}$ left-right-asymmetric 
rotation $\hat{K}_\mathrm{S}$ (ignoring the inner automorphism $\hat\Theta$), see 
section~\ref{sec:Z3AsymmetricRotation}.

\item On the circle of radius $1$, centered at the origin we obtain two left-right-asymmetric 
reflections $\hat{A}_\mathrm{refl.1} := (\hat{K}_\mathrm{S})^3 \hat{K}_*$ and 
$\hat{A}_\mathrm{refl.1'} := \hat{K}_\mathrm{S} \hat{K}_*$ corresponding to $v=\pm 1$ and $w=0$, 
respectively, see section~\ref{sec:Z3AsymmetricReflection}.
\end{enumerate}
Then, using the six-dimensional representations of $\Delta(54)$, $\hat{K}_*$, and 
$\hat{K}_\mathrm{S}$ from tables~\ref{tab:ModularTrafoTwistedStrings} and~\ref{tab:Delta54Bulk} one 
finds the unified flavor symmetry at $b=0$ and $r=\nicefrac{2}{\sqrt{3}}$ as the closure of all 
above transformations 1.-4., resulting in $\mathrm{SG}(216, 87)$. This is the unified flavor 
symmetry at all blue squares in figure~\ref{fig:modular_space1}. We stress that the specific 
left-right asymmetric rotations, 3., are already contained in the closure of 1.,2., and 4. That is, 
the symmetry at the intersecting points of lines and circles is already fully described by the 
intersecting symmetries and not enhanced beyond that.

\subsection[Unified flavor symmetry at b=1/2 and r=1: SG(324,39)]{\boldmath Unified flavor symmetry at $b=\nicefrac{1}{2}$ and $r=1$: $\mathrm{SG}(324,39)$\unboldmath}
\label{sec:SG324}

Next, we analyze the point $(b,r)=(\nicefrac{1}{2},1)$ in $T$-moduli space, 
figure~\ref{fig:modular_space2}. There, we identify the following type (1.) and type (2.b) outer 
automorphisms of the Narain space group:
\begin{enumerate}
\item At a generic point $(b,r)$ we found $\Delta(54)$ generated by $\mathrm{A}$, $\mathrm{B}$, 
and $\mathrm{C}$ as the traditional flavor symmetry, see section~\ref{sec:GenericPoint}.

\item On the red line at $b=\nicefrac{1}{2}$ we get additionally a left-right-symmetric reflection 
$\mathrm{D}(1) = (\hat{S}_\mathrm{refl.}(1), 0)$, where 
$\hat{S}_\mathrm{refl.}(1) = \hat{K}_\mathrm{T}\, \hat{K}_*$, see 
section~\ref{sec:Z3SymmetricReflection} for $n_B=1$.

\item Moreover, at $(b,r)=(\nicefrac{1}{2},1)$ there is a left-right-asymmetric $\Z{3}$ 
rotation $\hat{A}_\mathrm{rot.}\!\left(\nicefrac{2\pi}{3},0\right) = \hat{K}_\mathrm{T}\,\hat{K}_\mathrm{S}\,\hat{S}_\mathrm{rot.}\!\left(\nicefrac{2\pi}{6}\right)$, see section~\ref{sec:Z3AsymmetricRotation}.

\item On the circle of radius $1$, centered at the origin we obtain two left-right-asymmetric 
reflections $\hat{A}_\mathrm{refl.1} := (\hat{K}_\mathrm{S})^3 \hat{K}_*$ and 
$\hat{A}_\mathrm{refl.1'} := \hat{K}_\mathrm{S} \hat{K}_*$ corresponding to $v=\pm 1$ and $w=0$, 
respectively, see section~\ref{sec:Z3AsymmetricReflection}.

\item Finally, on the circle of radius $1$, centered at $(1,0)$ there are two left-right-asymmetric 
reflections $\hat{A}_\mathrm{refl.2} := \hat{K}_\mathrm{S} \hat{K}_* \hat{K}_\mathrm{T} \hat{K}_\mathrm{S}$ and 
$\hat{A}_\mathrm{refl.2'} := (\hat{K}_\mathrm{S})^3 \hat{K}_* \hat{K}_\mathrm{T} \hat{K}_\mathrm{S}$ corresponding to $v=w=\mp 1$, 
respectively, see section~\ref{sec:Z3AsymmetricReflection}.

\end{enumerate}
These automorphisms are not independent. Indeed, we identify the following relations
\begin{subequations}
\begin{eqnarray}
\hat{A}_\mathrm{refl.1'}                                 & = & \hat{S}_\mathrm{rot.}(\pi)\, \hat{A}_\mathrm{refl.1}\;, \\
\hat{A}_\mathrm{rot.}\!\left(\nicefrac{2\pi}{3},0\right) & = & \hat{S}_\mathrm{refl.}(1)\, \hat{A}_\mathrm{refl.1'}\, \hat\Theta^2\;, \\  
\hat{A}_\mathrm{refl.2}                                  & = & \hat{A}_\mathrm{refl.1}\, \hat{A}_\mathrm{rot.}\!\left(\nicefrac{2\pi}{3},0\right)\,\hat\Theta\;, \\
\hat{A}_\mathrm{refl.2'}                                 & = & \hat{S}_\mathrm{rot.}(\pi)\,\hat{A}_\mathrm{refl.2}\;,
\end{eqnarray}
\end{subequations}
where $\hat\Theta$ is an inner automorphism of the Narain space group. Consequently, the unified 
flavor symmetry at $(b,r)=(\nicefrac{1}{2},1)$ can be generated by the transformations $\mathrm{A}$, 
$\mathrm{B}$, and $\mathrm{C}$ from $\Delta(54)$ and by $\hat{S}_\mathrm{refl.}(1)$ and 
$\hat{A}_\mathrm{refl.1}$. Then, one can use tables~\ref{tab:ModularTrafoTwistedStrings} 
and~\ref{tab:Delta54Bulk} to obtain the six-dimensional representations of $\Delta(54)$ and of the 
generators
\begin{equation}
\hat{S}_\mathrm{refl.}(1) ~=~ \hat{K}_\mathrm{T}\, \hat{K}_* \quad\mathrm{and}\quad \hat{A}_\mathrm{refl.1} ~=~ \left(\hat{K}_\mathrm{S}\right)^3\,\hat{K}_*\;.
\end{equation}
Again using GAP to compute the closure, we find that this six-dimensional representation generates 
$\mathrm{SG}(324, 39)$. This is the unified flavor symmetry at $(b,r)=(\nicefrac{1}{2},1)$ and at 
all green curls in figure~\ref{fig:modular_space1}. Again, no extra generators besides those 
already conserved on the lines and semi-circles are needed. Finally, let us remark that at the 
green curls in figure~\ref{fig:modular_space1} also the gauge symmetry gets enhanced by a 
$\U{1}\times\U{1}$ factor, see e.g. ref.~\cite{Beye:2014nxa}.

%%%%%%%%%%%%%%%%%%%%%%%%%%%%%%%%%%%%%%%%%%%%%%%%%%%%%%%%%%%%%%%%%%%%%%%%%%%%%%%%%%%%%%%%%%%%%%%%%%%%%%%%%%%%%%%%%%%%%%%%%%%%%%%%%%%%%%%%%%
\section{Conclusions and outlook}
\label{sec:conclusions}

In the present paper we have presented a general method to deduce the flavor symmetries of string 
models. This led to a hybrid system of a unified flavor group composed of two distinct components. 
There is on one hand the traditional flavor group that is universal in moduli space. At some 
specific regions in moduli space, on the other hand, it is enhanced via duality symmetries that 
also include \CP-like transformations. The full flavor group is thus non-universal in moduli 
space and it allows different flavor- and \CP-structures for different sectors of the theory 
(dependent of the location of fields in the compact extra dimensions). At a generic point in moduli 
space the enhanced symmetries are broken spontaneously. For values of the moduli close to the 
self-dual points a hierarchy of flavor parameters can emerge. String theory thus provides us with 
some specific rules or lessons for flavor model building.

Up to now there has been substantial work on ``bottom-up'' model constructions of flavor that 
consider the concept of modular symmetries
~\cite{Altarelli:2005yx,    %hep-ph/0512103
deAdelhartToorop:2011re,    %1112.1340
Feruglio:2017spp,           %1706.08749
Kobayashi:2018vbk,          %1803.10391
Kobayashi:2018rad,          %1804.06644
Penedo:2018nmg,             %1806.11040
Criado:2018thu,             %1807.01125
Kobayashi:2018scp,          %1808.03012
Novichkov:2018ovf,          %1811.04933
Kobayashi:2018bff,          %1811.11384
Novichkov:2018nkm,          %1812.02158
deAnda:2018ecu,             %1812.05620
Okada:2018yrn,              %1812.09677
Kobayashi:2018wkl,          %1812.11072
Novichkov:2018yse,          %1812.11289
Ding:2019xna,               %1903.12588
Nomura:2019jxj,             %1904.03937
Kariyazono:2019ehj,         %1904.07546
Okada:2019uoy,              %1905.13421
deMedeirosVarzielas:2019cyj,%1906.02208
Nomura:2019yft,             %1906.03927
Kobayashi:2019rzp,          %1906.10341
Okada:2019xqk,              %1907.04716
Kobayashi:2019mna,           %1907.09141
Ding:2019zxk,
Okada:2019mjf}. 
The main focus there was on the description of the flavor structure of the lepton sector of the 
Standard Model (based on the finite modular groups $\Gamma_N$ for $N=2,3,4,5$), where modular 
transformations seem to be particularly successful. There has been less work on the quark 
sector~\cite{Okada:2018yrn,Kobayashi:2018wkl,Kobayashi:2019rzp,Okada:2019uoy} and the question of 
\CP symmetries has usually not been discussed (with the recent exception 
of~\cite{Novichkov:2019sqv}). Time has come to analyze possible connections of the bottom-up 
constructions with the rules and lessons from string theory presented in this paper. In order to 
compare the bottom-up constructions with the top-down picture, however, we first have to clarify 
some apparent differences between the two approaches:
\begin{itemize}
\item In string theory the modular transformations act nontrivially on the super- and 
K\"ahler-potential of the low-energy effective field theory~\cite{Ferrara:1989bc,Ferrara:1989qb}, 
while the phenomenological bottom-up models assume an invariant superpotential. It remains to be 
seen how this property of string theory can be accommodated in the bottom-up approach and whether 
(and how) this might affect the phenomenological predictions of the models.
\item String theory provides a hybrid flavor picture including the traditional flavor symmetries 
and \textit{parts} of the finite modular group $\Gamma_N$ (here $\Gamma_3$, and actually, its 
$\CP$-enhanced double covering group $\mathrm{GL}(2,3)$). That is, \textit{not} the full 
modular group may be realized in the low-energy effective theory: the stabilization of the 
$T$-modulus necessarily leads to a spontaneous breaking of parts of the modular group. In our 
example, the maximal enhancement of the \textit{non-modular} flavor symmetry by modular 
transformations is from $\Delta(54)$ to $\mathrm{SG}(324,39)$. This group only includes 
\textit{some} generators of $\mathrm{GL}(2,3)$. 
At this point we need more general string theory constructions to see whether present bottom-up constructions can be 
embedded in a string theory framework. In any case, we would expect that both, the traditional 
flavor symmetries and modular symmetries should be part of the fully unified picture of flavor and \CP.
\end{itemize}
More work is needed to answer these questions. This would require more explicit model building in 
string theory along the lines discussed in~\cite{Lebedev:2006kn,Lebedev:2007hv,Lebedev:2008un, Nilles:2011aj,Nilles:2014owa}.
One should keep in mind that our present discussion has concentrated on general aspects of the 
flavor structure, illustrated on a toy model in  $D=2$ compact extra dimensions. Even in this 
simplified case we were told a first lesson: string theory gives rise to potentially large flavor 
groups. In the simple $D=2$ example we already obtained a group as large as $\mathrm{SG}(324,39)$. 
This has to be generalized to $D=6$ within models that accommodate the spectrum of the standard 
model~\cite{Carballo-Perez:2016ooy}, where even larger groups are likely to 
emerge~\cite{Ramos-Sanchez:2017lmj,Olguin-Trejo:2018wpw}. With the tools described in the present 
paper we could then explore the full ``landscape'' of flavor symmetries in $D=6$, try to make 
connections to the existing bottom-up constructions, and extend the existing constructions of the 
lepton sector to a fully unified picture. A second generic lesson from string theory concerns \CP. 
It naturally appears as part of the modular symmetries at some specific regions in moduli space and 
is spontaneously broken if one moves away form these self-dual points and lines. The phenomenological 
properties will coincide with those discussed in ref.~\cite{Nilles:2018wex}, where \CP-violation is 
connected to the heavy winding modes of string theory. A third lesson from string theory is the 
appearance of a unified symmetry of flavor and \CP that is non-universal in moduli space. It 
includes the traditional flavor symmetry and modular symmetries at some specific regions on moduli 
space. Different sectors of a theory might have different flavor and \CP symmetries, and this might 
explain the different flavor and \CP structure of the quark and lepton sectors of the standard 
model. Reminiscent of the concept of local grand unification~\cite{Forste:2004ie,Buchmuller:2004hv} 
one might call this ``Local Flavor Unification'', as the flavor properties are connected to the 
location of fields in the compact extra dimensions. This provides a new perspective for flavor 
model building inspired by string theory.

\section*{Acknowledgments}

AB and PV are supported by the Deutsche Forschungsgemeinschaft (SFB1258). HPN thanks the CERN 
Theory Department for hospitality and support. PV would like to thank the Bethe Center for 
Theoretical Physics in Bonn for hospitality and support. AT would like to thank the Physics 
Department of TUM for hospitality during parts of this work.

%%%%%%%%%%%%%%%%%%%%%%%%%%%%%%%%%%%%%%%%%%%%%%%%%%%%%%%%%%%%%%%%%%%%%%%%%%%%%%%%%%%%%%%%%%%%%%%%%%%%%%%%%%%%%%%%%%%%%%%%%%%%%%%%%%%%%%%%%%
\appendix

\section[String states in Narain orbifolds]{\boldmath String states in Narain orbifolds \unboldmath}
\label{app:StringStates}

This appendix gives a brief review on string states on Narain orbifolds and their properties. We 
start by defining closed strings on Narain orbifolds via orbifold boundary conditions in 
section~\ref{app:BoundaryConditions}. Next, we analyze their transformation properties in 
section~\ref{app:TransformationsOfClosedStrings}, resulting in the observation that the outer 
automorphisms of the Narain space group $S_{\mathrm{Narain}}$ give rise to (flavor) symmetries of 
the string setup. Afterwards, section~\ref{sec:BulkVertexOperators} introduces vertex operators of 
untwisted strings, while section~\ref{sec:Z3VertexOperators} specializes to the example of the 
symmetric $\Z{3}$ orbifold in $D=2$ dimensions. Finally, in section~\ref{sec:TprimeDecomposition} 
some details on the irreducible representations of the finite modular group $\mathrm{T}'$ are 
presented with a focus on the transformation of the $\Delta(54)$ triplet of twisted strings under 
$\mathrm{T}'$ as $\rep{2}' \oplus \rep{1}$.

\subsection{Boundary conditions of closed strings}
\label{app:BoundaryConditions}

The orbifold boundary condition of a closed bosonic string $Y(\tau, \sigma)$ in the Narain 
formulation is given by
\begin{equation}\label{eq:NarainOrbifoldBC}
Y(\tau, \sigma+1) ~=~ g\, Y(\tau, \sigma) 
                  ~=~ \Theta^k\,Y(\tau, \sigma) + E\,\hat{N}\;, 
\end{equation}
where $g = (\Theta^k, E\,\hat{N}) \in S_{\mathrm{Narain}}$ is called the constructing element, see 
eq.~\eqref{eq:NarainOrbifold}. Since $Y \sim \tilde{g}\, Y$ are identified on the orbifold for all 
$\tilde{g} \in S_{\mathrm{Narain}}$ the boundary conditions with constructing elements $g$ and 
$\tilde{g}^{-1}g\,\tilde{g}$ describe the same closed string. Hence, a closed string on the 
orbifold is associated to a conjugacy class of constructing elements
\begin{equation}\label{eq:NarainCC}
[g] ~=~ \{ \tilde{g}^{-1}g\,\tilde{g} ~|~ \tilde{g}~\in~ S_{\mathrm{Narain}} \}\;.
\end{equation}
The resulting string state is denoted by $\ket{[g]}$.

Note that there is a transformation $y_\mathrm{R} \mapsto y_\mathrm{R} + \xi$ and 
$y_\mathrm{L} \mapsto y_\mathrm{L} - \xi$ such that 
$y \sim y_\mathrm{R} + y_\mathrm{L}$ is invariant~\cite{GrootNibbelink:2017usl}. This left-right 
asymmetric translation can be used match the Narain conjugacy class $[g]$ from 
eq.~\eqref{eq:NarainCC} to the corresponding conjugacy class of the geometrical space group.

\subsection{Transformations of closed strings under outer automorphisms}
\label{app:TransformationsOfClosedStrings}

A transformation with $h = (\Sigma, E\,\hat{T}) \not\in S_{\mathrm{Narain}}$ acts as 
\begin{equation}\label{eq:GeneralTrafoOfY}
Y ~\mapsto~ h\,Y ~=~ \Sigma\,Y + E\,\hat{T}\;.
\end{equation}
Consequently, $h$ transforms a boundary condition~\eqref{eq:NarainOrbifoldBC} with constructing 
element $g \in S_{\mathrm{Narain}}$ according to
\begin{equation}
h\, Y(\tau, \sigma+1) ~=~ g\, h\, Y(\tau, \sigma) \qquad \Leftrightarrow \qquad Y(\tau, \sigma+1) ~=~ \left(h^{-1}g\, h\right)\, Y(\tau, \sigma)\;.
\end{equation}
To ensure that $h$ is a consistent transformation, this boundary condition must belong to some, maybe 
different, constructing element of the orbifold theory. Hence,
\begin{equation}
h^{-1}g\, h ~\in~ S_{\mathrm{Narain}} \quad\mathrm{even\ though}\quad h \not\in S_{\mathrm{Narain}}\;,
\end{equation}
for all $g \in S_{\mathrm{Narain}}$. Consequently, a transformation $h\not\in S_{\mathrm{Narain}}$ 
must be an outer automorphism of the Narain space group $S_{\mathrm{Narain}}$ and, in general, it 
acts nontrivially on string states, i.e.\
\begin{equation}\label{eq:TrafoOfStringState}
\ket{[g]} ~\stackrel{h}{\mapsto}~ \varphi_{g,h}\,\ket{[h^{-1}g\, h]}\;,
\end{equation}
up to a possible phase $\varphi_{g,h}$.

\paragraph{Examples:} Let us discuss two examples. First we take an untwisted string with 
constructing element $g = (\Id, E\,\hat{N})\in S_{\mathrm{Narain}}$, i.e.\ a string with winding 
and KK numbers given by $\hat{N} \in \Z{}^{2D}$ that lives in the bulk of the orbifold. Then, we 
analyze the action of a purely rotational outer automorphism 
$h = (\Sigma,0)\not\in S_{\mathrm{Narain}}$. Using $\hat{\Sigma} := E^{-1}\Sigma\,E$ in 
eq.~\eqref{eq:TrafoOfStringState} yields
\begin{equation}
\ket{[(\Id, E\,\hat{N})]} ~\stackrel{h}{\mapsto}~ \ket{[(\Id, \Sigma^{-1}\,E\,\hat{N})]} ~=~ \ket{[(\Id, E\,\hat{\Sigma}^{-1}\,\hat{N})]} \;,
\end{equation}
where we have already used that the phase $\varphi_{g,h}$ of an untwisted string $g$ is trivial for 
a purely rotational transformation $h$, as we will see explicitly in eq.~\eqref{eqn:TrafoOfBulkVertexOperator} 
in appendix~\ref{sec:BulkVertexOperators}.

As a second example, take a twisted string with constructing element 
$g = (\Theta, E\,\hat{N})\in S_{\mathrm{Narain}}$ which is localized at the fixed point of $g$. In 
this case, we take an outer automorphism $h = (\Id, E\,\hat{T})\not\in S_{\mathrm{Narain}}$ and 
eq.~\eqref{eq:TrafoOfStringState} yields
\begin{equation}
\ket{[(\Theta, E\,\hat{N})]} ~\stackrel{h}{\mapsto}~ \varphi_{g,h}\, \ket{[(\Theta, E\,\hat{N} - (\Id-\Theta)\, E\,\hat{T})]}\;.
\end{equation}
Hence, translations $h = (\Id, E\,\hat{T})\not\in S_{\mathrm{Narain}}$ can permute twisted string 
states localized at different fixed points. The determination of the phase $\varphi_{g,h}$ of a 
twisted string is more involved. An example is given in section~\ref{sec:GenericPoint}.

\subsection[Untwisted vertex operators]{Untwisted vertex operators}
\label{sec:BulkVertexOperators}

A closed bosonic string compactified on a $D$-dimensional torus with winding numbers $n\in\Z{}^D$ 
and KK numbers $m\in\Z{}^D$ corresponds to a string eq.~\eqref{eq:NarainOrbifoldBC} with 
constructing element $g=(\Id, E\,\hat{N})\in S_{\mathrm{Narain}}$ and the associated vertex operator 
reads
\begin{equation}\label{eqn:StringVertexOperatorForBulk}
V(\hat{N}) ~=~ \exp\big(2\pi \I P^\mathrm{T} \eta\, Y\big) \quad\mathrm{where}\quad P ~=~ \left(\!\begin{array}{c}p_\mathrm{R}\\p_\mathrm{L}\end{array}\!\right) ~=~ E\, \hat{N} \quad\mathrm{and}\quad \hat{N} ~=~ \left(\!\begin{array}{c}n\\m\end{array}\!\right) ~\in~\Z{}^{2D}\;,
\end{equation}
ignoring the co-cycle and normal-ordering. The Narain momentum $P=E\hat{N}$ contains right- and 
left-moving momenta that enter the string's total mass, i.e.\ 
\begin{equation}\label{eq:UntwistedStringMass}
M^2 ~\propto~ P^2 ~=~ (p_\mathrm{R})^2 + (p_\mathrm{L})^2 ~=~ \hat{N}^\mathrm{T} \mathcal{H} \hat{N}\;,
\end{equation}
plus further contributions and using $\mathcal{H} = E^\mathrm{T} E$. Compared to 
refs.~\cite{Lauer:1990tm,Nilles:2018wex} we have set $V(\hat{N})\equiv V^{p,w}$, where the momentum 
vector $p$ and the winding vector $w$ are given by the KK numbers $m$ and winding numbers $n$, 
respectively. 

Let us analyze the transformation of a vertex operator $V(\hat{N})$ under a general transformation 
$h=(\Sigma, E\,\hat{T})$ of the right- and left-moving bosonic string coordinates 
$Y \stackrel{h}{\mapsto} \Sigma\,Y + E\,\hat{T}$ from eq.~\eqref{eq:GeneralTrafoOfY}. This yields
\begin{equation}\label{eqn:TrafoOfBulkVertexOperator}
V(\hat{N}) ~\stackrel{h}{\mapsto}~ \exp\left(2\pi\I \hat{N}^\mathrm{T}\hat\eta\,\hat{T}\right)\,V(\hat\Sigma^{-1}\,\hat{N})\;,
\end{equation}
under the assumption $\hat\Sigma = E^{-1} \Sigma\,E \in \mathrm{O}_{\hat{\eta}}(D,D,\Z{})$, see 
eq.~\eqref{eqn:SymmetryGroupOfNarainLattice}. We will use this result frequently, when we discuss 
the transformation of bosonic strings under outer automorphisms of the $\Z{3}$ Narain space group, 
see for example section~\ref{sec:Z3ModularTransformationsOfStrings}. Furthermore, note that this 
transformation eq.~\eqref{eqn:TrafoOfBulkVertexOperator} is in agreement with 
$\hat{N} \mapsto \hat{N}' = \hat\Sigma^{-1} \hat{N}$ from eq.~\eqref{eqn:SymmetryTrafoOfNarainLattice}.

Using eq.~\eqref{eqn:TrafoOfBulkVertexOperator}, one easily verifies that a vertex operator 
$V(\hat{N})$ is invariant under a shift by a Narain lattice vector $Y \mapsto Y + E\,\hat{T}$ with 
$\hat{T}\in\Z{}^{2D}$, i.e.
\begin{equation}
V(\hat{N}) ~\mapsto~ \underbrace{\exp\big(2\pi \I \hat{N}^\mathrm{T} \hat\eta\,\hat{T}\big)}_{=~1}\, V(\hat{N}) ~=~ V(\hat{N})\;.
\end{equation}
On the other hand, for a fractional shift $Y \mapsto Y + E\,\hat{T}$ with $\hat{T}\not\in\Z{}^{2D}$ 
a vertex operator $V(\hat{N})$ obtains in general a nontrivial phase. 

Under the $\Z{K}$ orbifold action $Y \mapsto \Theta\,Y$ with $\Theta\, E = E\,\hat\Theta$ a bosonic 
string vertex operator eq.~\eqref{eqn:StringVertexOperatorForBulk} transforms as 
$V(\hat{N}) ~\mapsto~ V(\hat\Theta^{-1}\,\hat{N})$\;, using eq.~\eqref{eqn:TrafoOfBulkVertexOperator}. 
Consequently, assuming orbifold-invariance of the other string degrees of freedom, the $\Z{K}$ 
orbifold-invariant combination for $\hat{N} \neq 0$ reads 
\begin{equation}\label{eqn:VertexOperatorZKOrbifoldInvariant}
V(\hat{N})^\mathrm{orb.} ~=~ \frac{1}{\sqrt{K}} \sum_{k=0}^{K-1}\, V(\hat\Theta^k\,\hat{N})\;,
\end{equation}
and the orthogonal linear combinations are removed from the orbifold spectrum. This vertex operator 
corresponds to an orbifold-invariant string state $\ket{[g]}$ with constructing element 
$g=(\Id,E\,\hat{N})\in S_{\mathrm{Narain}}$.

\subsection[Vertex operators of the Z3 Narain orbifold]{\boldmath Vertex operators of the $\Z{3}$ Narain orbifold\unboldmath}
\label{sec:Z3VertexOperators}

Let us now specialize to the symmetric $\Z{3}$ orbifold in two dimensions, see 
section~\ref{sec:Z3NarainSpaceGroup}. The full particle spectrum of the $\Z{3}$ Narain orbifold 
contains untwisted strings with constructing elements $g=(\Id, E\,\hat{N})\in S_{\mathrm{Narain}}$, 
where $\hat{N} \in \Z{}^4$ gives the winding numbers $n$ and KK numbers $m$. Then, the 
orbifold-invariant untwisted vertex operators read
\begin{equation}\label{eq:Z3NarainBulkVertexOperator}
V(\hat{N})^\mathrm{orb.} ~=~ \frac{1}{\sqrt{3}}\left( V(\hat{N}) + V(\hat\Theta\,\hat{N}) + V(\hat\Theta^2\,\hat{N}) \right) \quad\mathrm{for}\quad \hat{N} ~=~ \left(\begin{array}{c}n\\m\end{array}\right) ~\in~\Z{}^{4}\;,
\end{equation}
for $\hat{N} \neq 0$ and $V(\hat{N})^\mathrm{orb.} = V(\hat{N})$ for $\hat{N} = 0$. In addition, 
there are twisted strings $(X,\,Y,\,Z)$ with constructing elements $(\Theta, E\,\hat{N})$ from the 
first twisted sector and $(\bar{X},\,\bar{Y},\,\bar{Z})$ with constructing elements 
$(\Theta^2, E\,\hat{N})$ from the second twisted sector, where we focus on the respective twist 
fields. Furthermore, we note that in the following we ignore other contributions to the full string 
vertex operators like co-cycles, world-sheet fermions, oscillator excitations, and the 16 gauge 
degrees of freedom. Their inclusion can only yield additional transformation phases but cannot 
change the non-Abelian structure of the flavor groups which is the main concern of this work.

For an untwisted string $V(\hat{N})^\mathrm{orb.}$ one can define two discrete $\Z{3}$ charges: KK 
charge $M \in\{0,1,2\}$ and winding charge $N\in\{0,1,2\}$, i.e.\
\begin{equation}
(M, N) ~=~ (-m_1 + m_2, n_1 + n_2)\;,
\end{equation}
where both charges are defined mod 3, see e.g.\ ref.~\cite{Nilles:2018wex}. Note that each term 
$V(\hat\Theta^k\,\hat{N})$ with $k=0,1,2$ in the orbifold-invariant vertex 
operator~\eqref{eq:Z3NarainBulkVertexOperator} carries the same $\Z{3}$ charges, i.e.\ 
\begin{equation}
(M, N) ~\stackrel{~\hat\Theta~}{\longmapsto}~ (2m_1 + m_2, n_1 - 2n_2) ~=~ (M, N)\;,
\end{equation}
using that $M$ and $N$ are defined modulo 3. Then, we can arrange all orbifold-invariant 
untwisted strings $V(\hat{N})^\mathrm{orb.}$ into nine classes $V^{(M,N)}$ depending on their 
$\Z{3}\times\Z{3}$ charges $(M,N)$ using $\hat{N} \in \Gamma^{(M,N)} \subset \Z{}^4$ for $M,N \in \{0,1,2\}$, 
where we define the sublattices
\begin{equation}
\Gamma^{(M,N)} ~=~ \{\hat{N} \in \Z{}^4 ~|~ \hat{N} ~\sim~ \hat\Theta\,\hat{N} \;\mathrm{and}\; M = -m_1 + m_2 \;\mathrm{ mod }\;3\;\mathrm{ and }\; N = n_1 + n_2 \;\mathrm{ mod }\;3\}\;.
\end{equation}
Explicitly, the nine classes $V^{(M,N)}$ of orbifold-invariant untwisted strings $V(\hat{N})^\mathrm{orb.}$ 
are defined as
\begin{equation}
V^{(M,N)} ~=~ \sum_{\hat{N} ~\in~ \Gamma^{(M,N)}} C(\hat{N})\, V(\hat{N})^\mathrm{orb.} \quad\mathrm{for}\quad M,N ~=~ 0,1,2\;,
\end{equation}
where the coefficients $C(\hat{N})$ are given in ref.~\cite{Lauer:1990tm}.

The classes $V^{(M,N)}$ of untwisted strings appear in the operator product expansions (OPEs) 
between twisted strings $(X,\,Y,\,Z)$ from the first twisted sector and twisted strings 
$(\bar{X},\,\bar{Y},\,\bar{Z})$ from the second twisted sector~\cite{Lauer:1990tm}, i.e.\
\begin{subequations}\label{eqn:InvertedOPEs}
\begin{eqnarray}
V^{(0,0)} & = & \frac{1}{3} \left(X\bar{X} + Y\bar{Y} + Z\bar{Z} \right)\;, \\ 
\left(\begin{array}{c}V^{(0,2)} \\V^{(0,1)} \end{array}\right) & = &
\frac{1}{3}\left(
\begin{array}{c}
Y\bar{Z} + Z\bar{X} + X\bar{Y}\\
Z\bar{Y} + X\bar{Z} + Y\bar{X}
\end{array}\right)\;,\\ 
\left(\begin{array}{c}V^{(1,0)} \\V^{(2,0)} \end{array}\right) & = &
\frac{1}{3}\left(\begin{array}{c}
X\bar{X} + \omega^{\phantom{2}}\, Y\bar{Y} + \omega^2\, Z\bar{Z} \\
X\bar{X} + \omega^2\, Y\bar{Y} + \omega^{\phantom{2}}\, Z\bar{Z}
\end{array}\right)\;,\\
\left(\begin{array}{c}V^{(1,2)} \\V^{(2,1)} \end{array}\right) & = &
\frac{1}{3}\left(\begin{array}{c}
Y\bar{Z} + \omega^{\phantom{2}}\, Z\bar{X} + \omega^2\, X\bar{Y}\\
Z\bar{Y} + \omega^2\, X\bar{Z} + \omega^{\phantom{2}}\, Y\bar{X}
\end{array}\right)\;,\\
\left(\begin{array}{c}V^{(1,1)} \\V^{(2,2)}\end{array}\right) & = &
\frac{1}{3}\left(\begin{array}{c}
Z\bar{Y} + \omega^{\phantom{2}}\, X\bar{Z} + \omega^2\, Y\bar{X}\\
Y\bar{Z} + \omega^2\, Z\bar{X} + \omega^{\phantom{2}}\, X\bar{Y}
\end{array}\right)\;,
\end{eqnarray}
\end{subequations}
where $\omega=\exp(\nicefrac{2\pi\I}{3})$. These OPEs will turn out to be crucial in order to 
translate the transformation properties eq.~\eqref{eqn:TrafoOfBulkVertexOperator} of untwisted 
strings to the twisted strings.

\subsection[Irreducible representations of the finite modular group T']{\boldmath Irreducible representations of the finite modular group $\mathrm{T}'$\unboldmath}
\label{sec:TprimeDecomposition}

\begin{table}[!t]
\begin{center}
\begin{tabular}{l|rrrrrrr}
\toprule
$\mathrm{T}'$ & $\big[\Id\big]$ & $\big[\mathrm{s}^2\big]$ & $\big[\mathrm{t}\big]$ & $\big[\mathrm{t}^2\big]$ & $\big[\mathrm{s}\big]$ & $\big[\mathrm{s^2t}\big]$ & $\big[\mathrm{s^2t^2}\big]$ \\\midrule
$\rep{1}$     & $1$ & $1$  & $1$                           & $1$      & $1$ & $1$ & $1$ \\
$\rep{1}'$    & $1$ & $1$  & $\omega$                      & $\omega^2\nphantom{${}^2$}$ & $1$ & $\omega$ & $\omega^2\nphantom{${}^2$}$ \\
$\rep{1}''$   & $1$ & $1$  & $\omega^{2}\nphantom{${}^2$}$ & $\omega$ & $1$ & $\omega^2\nphantom{${}^2$}$ & $\omega$ \\
$\rep{2}$     & $2$ & $-2$ & $-1$                          & $-1$ & $0$ & $1$ & $1$ \\
$\rep{2}'$    & $2$ & $-2$ & $-\omega$                     & $-\omega^2\nphantom{${}^2$}$ & $0$ & $\omega$ & $\omega^2\nphantom{${}^2$}$ \\
$\rep{2}''$   & $2$ & $-2$ & $-\omega^{2}\nphantom{${}^2$}$& $-\omega$ & $0$ & $\omega^2\nphantom{${}^2$}$ & $\omega$ \\
$\rep{3}$     & $3$ & $3$  & $0$                           & $0$ & $-1$ & $0$ & $0$ \\
\bottomrule		
\end{tabular}
\caption{Character table of the finite modular group $\mathrm{T}' \cong \mathrm{SL}(2,3)$. Here, 
$\omega:=\mathrm{e}^{\frac{2\pi\I}{3}}$.}
\label{tab:CharacterTableTprime}
\end{center}
\end{table}

In addition to their transformation under the traditional flavor symmetry $\Delta(54)$, where they 
transform as $\rep{6}=\rep{3} \oplus \crep{3}$, the twisted string states 
$(X,\,Y,\,Z,\,\bar{X},\,\bar{Y},\,\bar{Z})$ of the $\Z{3}$ orbifold also transform under modular 
transformations $\mathrm{S}, \mathrm{T} \in \mathrm{SL}(2,\Z{})_T$ with the six-dimensional 
matrices $\hat{K}_{\mathrm{S}, \rep{6}}$ and $\hat{K}_{\mathrm{T}, \rep{6}}$ given in 
table~\ref{tab:ModularTrafoTwistedStrings}. However, these two matrices do not correspond to a 
faithful representation of $\mathrm{SL}(2,\Z{})_T$ (for example, 
$(\hat{K}_{\mathrm{T}, \rep{6}})^3=\Id$ even though $\mathrm{T} \in \mathrm{SL}(2,\Z{})_T$ has 
infinite order) but they generate the finite modular group $\mathrm{T}' \cong \mathrm{SL}(2,3)$ 
(the double covering group of $A_4 \cong \Gamma_3$ of order 24), which can be defined by the 
presentation\footnote{%	
An in depth discussion of the group $\mathrm{T}'$ is given in chapter 5 of ref.~\cite{Ishimori:2010au}. 
The generators used there are related to our generators as $s\mathrel{\hat=}(\hat{K}_{\mathrm{S}})^3$ 
and $t\mathrel{\hat=}\hat{K}_{\mathrm{T}}$.}
\begin{equation}
\mathrm{T}' ~=~ \left\langle \mathrm{s},\,\mathrm{t} ~\Big|~  
\mathrm{s}^4 \,=\, \mathrm{t}^3 \,=\, (\mathrm{s}\,\mathrm{t})^3 \,=\, \Id \,,~\mathrm{s}^2\,\mathrm{t} \,=\, \mathrm{t}\,\mathrm{s}^2 \right\rangle\;.
\end{equation}

Indeed, the matrices $\hat{K}_{\mathrm{S}, \rep{6}}$ and $\hat{K}_{\mathrm{T}, \rep{6}}$ generate a 
\textit{reducible}, six-dimensional representation of $\mathrm{T}'$. They are block-diagonal with 
$3\times3$ blocks corresponding to the strings from first and second twisted sector, respectively 
(these blocks are exchanged by the action of the \CP-like transformation $\hat{K}_*$, as expected). 
This six-dimensional representation decomposes into irreducible representations of $\mathrm{T}'$ as 
(see also refs.~\cite{Lerche:1989cs,Kobayashi:2018rad}
\begin{equation}\label{eq:TprimeDecomposition}
\rep{6} ~=~ \left(\rep{2}' \oplus \rep{1}\right) \oplus \left(\rep{2}'' \oplus \rep{1}\right)\;.
\end{equation}
This decomposition can be made explicit by the following basis change: Focusing on the upper 
three-dimensional block only, $(X,\,Y,\,Z)$ are rotated into $X_0=-X$ and 
$X_{\pm}=(Y\pm Z)/\sqrt{2}$ by the orthogonal transformation
\begin{equation}
\begin{pmatrix}X_+\\X_0\\X_-\end{pmatrix}
~=~
\begin{pmatrix}0&\frac{1}{\sqrt{2}}&\frac{1}{\sqrt{2}}\\-1&0&0\\0&\frac{1}{\sqrt{2}}&-\frac{1}{\sqrt{2}}\end{pmatrix}
\begin{pmatrix}X\\Y\\Z\end{pmatrix}\;.
\end{equation}
Then, for the first twisted sector, $\hat{K}_{\mathrm{S}}$ and $\hat{K}_{\mathrm{T}}$ take the form
\begin{equation}
\hat{K}_{\mathrm{S}, \rep{3}}' ~=~ \left(\begin{array}{cc|c}\nicefrac{\I}{\sqrt{3}}&\sqrt{\nicefrac{2}{3}}\,\I&0\\\sqrt{\nicefrac{2}{3}}\,\I&-\nicefrac{\I}{\sqrt{3}}&0\\\hline 0&0&1\end{array}\right)\qquad\mathrm{and}\qquad
\hat{K}_{\mathrm{T}, \rep{3}}' ~=~ \left(\begin{array}{cc|c}1&0&0\\0&\omega^2&0\\\hline 0&0&1\end{array}\right)\;,
\end{equation}
proving the $2\oplus 1$ block-structure. The basis change for the lower $3\times3$ blocks of 
$\hat{K}_{\mathrm{S}, \rep{6}}$ and $\hat{K}_{\mathrm{T}, \rep{6}}$ works completely analogous, and 
the second twisted sector states $(\bar{X},\,\bar{Y},\,\bar{Z})$ transform with the complex 
conjugate of the above matrices. Using the character table of $\mathrm{T}'$, as given in 
table~\ref{tab:CharacterTableTprime}, it is straightforward to verify the 
decomposition~\eqref{eq:TprimeDecomposition}. In summary, the three twisted strings $(X,\,Y,\,Z)$ 
corresponding to the three fixed points of the two-dimensional $\Z{3}$ orbifold \textit{do not} 
transform as an irreducible $\rep{3}$ of $\mathrm{T}'$ but as a doublet $\rep{2}'$ and a trivial 
singlet $\rep{1}$.

If one additionally takes $\hat{K}_*$ into account, $\mathrm{T}' \cong \mathrm{SL}(2,3)$ gets 
enlarged to $\mathrm{GL}(2,3)$ and the six-dimensional representation 
eq.~\eqref{eq:TprimeDecomposition} decomposes into irreducible representations of $\mathrm{GL}(2,3)$ 
as $\rep{4}\oplus\rep2$.

\bibliography{Orbifold}
\bibliographystyle{OurBibTeX}
\end{document}